\documentclass[11pt]{article}
\usepackage{jheppub}
\usepackage{epsfig} 
\usepackage{amssymb}
\usepackage{amsmath}
\allowdisplaybreaks[4]
\usepackage{epstopdf}
\usepackage{extarrows}
\usepackage{verbatim}
\usepackage{dsfont,bbm}
\usepackage{subfig}

\usepackage{shuffle}
\usepackage{booktabs}
\usepackage{array}
\usepackage{tikz}
\usetikzlibrary{positioning, shapes.geometric}

\usepackage[textsize=footnotesize,textwidth=2.25cm]{todonotes}
\graphicspath{{./figs/}}
\allowdisplaybreaks[4]

\newcommand{\itbf}[1]{\textbf{\textit{#1}}}
\newcommand{\dt}[1]{\Delta_{#1}^{[2]}}

\title{Applying Color-Kinematics Duality in Pure Yang-Mills at Three Loops}

\author[a,b]{Zeyu Li,}
\emailAdd{lizeyu@itp.ac.cn}
\author[a,b,c,d]{Gang Yang,}
\emailAdd{yangg@itp.ac.cn}
\author[a,b]{and Guorui Zhu}
\emailAdd{zhuguorui@itp.ac.cn}
\affiliation[a]{CAS Key Laboratory of Theoretical Physics, Institute of Theoretical Physics, \\Chinese Academy of Sciences, Beijing 100190, China}
\affiliation[b]{School of Physical Sciences, University of Chinese Academy of Sciences, Beijing 100049, China}
\affiliation[c]{School of Fundamental Physics and Mathematical Sciences, Hangzhou Institute for Advanced Study, UCAS, Hangzhou 310024, China}
\affiliation[d]{International Centre for Theoretical Physics Asia-Pacific, Beijing/Hangzhou, China}

\abstract{
We present the first application of color-kinematics (CK) duality at the three-loop level in non-supersymmetric pure Yang-Mills (YM) theory. Building on the minimal deformation approach introduced in \cite{Li:2023akg}, we extend its use to the three-loop Sudakov form factor. Although three classes of unitarity cuts fail under the globally off-shell CK-dual ansatz, a compact and elegant solution is achieved by deforming a single master numerator. The final numerators exhibit Lorentz invariance in $d$ dimensions and take a local form. This method harnesses CK duality's full potential by enforcing a subset of off-shell dual Jacobi identities for the deformation, offering a promising path toward constructing three-loop amplitudes in non-supersymmetric YM theory and gravity through CK duality and double copy.
}

\begin{document}

\maketitle

\setcounter{footnote}{0}

\section{Introduction}

The conjectured color-kinematics (CK) duality \cite{Bern:2008qj, Bern:2010ue} suggests a profound connection between the kinematic and color structures in gauge theories, with significant implications for our understanding of both gauge and gravitational theories. 
This duality offers a powerful framework for constructing full-color gauge-theory quantities by linking the planar and non-planar components through dual kinematic relations. Additionally, it facilitates the construction of gravitational amplitudes directly from gauge-theory amplitudes, provided that the latter are organized to respect the CK duality.

At tree level, the CK duality has been established using both string theory and gauge theory methods \cite{BjerrumBohr:2009rd, Stieberger:2009hq, Feng:2010my, Bjerrum-Bohr:2010pnr, Mafra:2011kj}. However, at loop level, the duality remains conjectural and has been verified only through explicit examples, including both amplitudes and form factors \cite{Bern:2010tq, Carrasco:2011mn,Bern:2012uf, Du:2012mt, Oxburgh:2012zr, Yuan:2012rg,Boels:2012ew, Boels:2013bi, Bjerrum-Bohr:2013iza, Bern:2013yya, Ochirov:2013xba, Mafra:2015mja,
He:2015wgf, Mogull:2015adi, Yang:2016ear, Chiodaroli:2017ngp,Boels:2017skl, He:2017spx,Johansson:2017bfl,Jurado:2017xut, Geyer:2017ela, Faller:2018vdz,
Lin:2020dyj,Edison:2020uzf,Lin:2021qol,Lin:2021lqo, Li:2022tir}.
For a comprehensive review of CK duality and its connection to the double-copy construction, see \cite{Bern:2019prr,Bern:2022wqg}. 

In supersymmetric theories, CK duality has been extended to relatively high loop levels. For example, explicit solutions that manifest CK duality have been found for the four-loop four-point amplitude \cite{Bern:2012uf} and the five-loop Sudakov form factor \cite{Yang:2016ear} in ${\cal N}=4$ SYM. 
On the other hand, in non-supersymmetric gauge theories, constructing CK-dual loop integrands has proven far more challenging, with only limited success at two loops. Examples include two-loop four-gluon and five-gluon amplitudes where all helicities are equal \cite{Bern:2013yya, Mogull:2015adi}. No results at the three-loop level are currently available.

A major obstacle in extending CK duality to high loops in pure Yang-Mills (YM) theory is the absence of simple, globally off-shell CK-dual solutions. For instance, it was shown in \cite{Bern:2015ooa} (and later confirmed in \cite{Li:2023akg, Edison:2023ulf}) that the two-loop four-gluon amplitude in pure YM cannot exhibit global CK duality in a Lorentz-invariant local form in $d$ dimensions.

To address this issue, a new strategy for implementing CK duality was recently proposed in \cite{Li:2023akg}. 
The core idea of \cite{Li:2023akg} involves introducing a small ``deformation" that allows the CK duality relations to be applied effectively. This approach was shown to work extremely well for the two-loop four-gluon amplitude in pure YM theory. 
In particular, the deformation satisfies a subset of off-shell dual Jacobi relations, enabling efficient use of CK duality while maintaining a compact ansatz. 
The resulting numerators, which are remarkably simple, are presented in $d$-dimensionally Lorentz invariant local form and valid for all helicities of external gluons.
These findings suggest that the global off-shell CK duality is only slightly violated.

In this paper, we explore the CK duality for the first time at a three-loop level in pure YM theory.
Specifically, we extend the deformation strategy of \cite{Li:2023akg} and apply it to the Sudakov form factor.\footnote{The three-loop Sudakov form factor in pure YM was obtained earlier using Feynman diagram methods \cite{Baikov:2009bg, Gehrmann:2010ue}, but our focus is on its novel property of the CK duality at the integrand level.}
The complexity of the three-loop form factor significantly exceeds that of the two-loop four-gluon amplitude. First, the number of topologies increases from 14 at two loops to 58 at three loops. Furthermore, for the global CK-dual ansatz, three types of unitarity cuts fail for the three-loop form factor, compared to just one for the two-loop amplitude. 
Despite these challenges, we demonstrate that the deformation strategy remains effective for the three-loop form factor. Notably, only a single master numerator requires deformation, and the final solution is presented in a highly compact form. 

This non-trivial three-loop result corroborates the picture that CK duality may be only mildly broken in general. 
Our investigation of the three-loop form factor lays the groundwork for future applications of CK duality in computing three- or higher-loop full-color gauge and gravitational amplitudes in non-supersymmetric theories.

This paper is structured as follows. 
In Section~\ref{sec:review}, we provide a brief review of CK duality and outline the construction procedure, including the unitarity method.
The main results of this paper are presented in Section~\ref{sec:ff3loop} and Section~\ref{sec:FFdeformation}.
In Section~\ref{sec:ff3loop}, we explore the global CK-dual relations for the three-loop Sudakov form factor and highlight the incompatibility between the minimal ansatz and the unitarity cuts. We then introduce the deformation and derive the physical solution in Section~\ref{sec:FFdeformation}.
In Section~\ref{sec:discussion}, we summarize the result and discuss the connection between the Sudakov form factor and the four-gluon amplitude.
Additionally, the numerator solution, along with the relevant propagator lists and CK relations, is provided in the ancillary files.

\section{Review}

\label{sec:review}
In this section, we review the basic concept of CK duality and the general strategy for
constructing CK-dual integrands at the loop level.

\subsection{Review of CK duality}
The CK duality  \cite{Bern:2008qj} conjectures that there exists a cubic graph representation of amplitudes in
which the kinematic numerators satisfy the same equations of Jacobi relations for the color
factors.

The most important and basic example that illustrates CK duality is the four-gluon tree amplitude shown in Figure~\ref{treeA4}:

\begin{figure}[t]
	\centerline{\includegraphics[height=2.5cm]{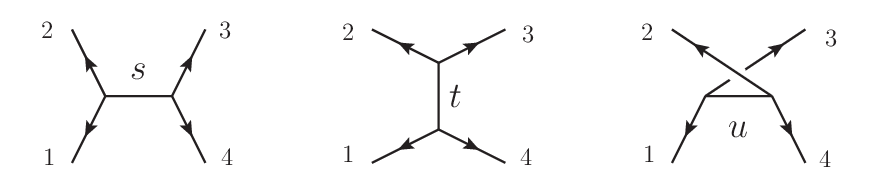} } 
	\caption{Trivalent graphs for the four-point tree amplitude.} 
	\label{treeA4}
\end{figure}

\begin{equation}\label{eq:fullcolortreeA4}
	\itbf{A}_4^{\rm tree}=g^2\left(\frac{c_sn_s}{s}+\frac{c_tn_t}{t}+\frac{c_un_u}{u}\right)\,,
\end{equation}
which is expanded in terms of the three trivalent topologies. The color factor is
defined as:
\begin{equation}\label{color_def}
	c_s=\tilde{f}^{a_1a_2s}\tilde{f}^{sa_3a_4},\qquad c_t=\tilde{f}^{a_2a_3t}\tilde{f}^{ta_4a_1}, \qquad c_u=\tilde{f}^{a_1a_3u}\tilde{f}^{ua_2a_4},
\end{equation}
with $\tilde{f}^{abc}\text{=i}\sqrt{2}f^{abc}\text{=tr([}T^a,T^b\text{],}T^c\text{)}$ being structure constant of $SU(N_c)$ gauge group. We abbreviate $\tilde{f}$ as $f$ in the rest of the paper.

The Jacobi relation of $f^{abc}$ gives
\begin{equation}\label{stu_Jacobi}
 	c_s= c_t + c_u \,,
\end{equation}
which comes directly from the definition of color factors.
The key property of CK duality is that the kinematic numerators $n_i$ should satisfy the same equation:
\begin{equation}
	n_s= n_t + n_u \,.
\end{equation}
We will refer to this relation as ``dual Jacobi relation'' or ``CK relation''.

In this simple case, it is easy to calculate $n_{s,t,u}$ using Feynman rules and check that they indeed satisfy the Jacobi relation. For higher point tree amplitude, it has been proven that the CK dual numerators can always be constructed \cite{BjerrumBohr:2009rd, Stieberger:2009hq, Feng:2010my, Bjerrum-Bohr:2010pnr, Mafra:2011kj}. However, at the loop level, the existence of CK dual numerators remains to be a conjecture in general and we need to check it case by case.

In this paper, we will primarily focus on the Sudakov form factor of the operator $\text{tr}(F^2)$, which is defined as
\begin{equation}
	\itbf{F}_2 = \int d^d x\ e^{-iq \cdot x} \langle g(p_1) g(p_2) | \text{tr}(F^2)(x) | 0 \rangle \,,
	\label{eq: defition-f2}
\end{equation}
where $ q = p_1  + p_2 $ is the off-shell momentum carried by the operator, and $p_i$ are the momenta of on-shell external gluons. At the tree level, the form factor is 
\begin{equation}
	\itbf{F}_2^{(0)} = C F_2^{(0)}\ .
	\label{eq:f2-minimal}
\end{equation}
Here we split it up into two parts. The first part $C = \text{tr}(T^{a_1} T^{a_2}) = \delta^{a_1 a_2 }$ is the color factor and the second part is called the color-stripped form factor with 
\begin{equation}
	F_2^{(0)} = (\varepsilon_1 \cdot \varepsilon_2)(p_1 \cdot p_2) - (\varepsilon_1 \cdot p_2)(\varepsilon_2 \cdot p_1) \,.
	\label{eq:f2-minimal-expr}
\end{equation}

The general form of $l$-loop Sudakov form factor can be given as
\begin{equation}
	\itbf{F}_2^{(l)}= \sum\limits_{\sigma_2} \sum\limits_{\Gamma_i}  \int \prod\limits_{j=1}^{l} \frac{d^{D}l_j}{(2\pi)^D} \frac{1}{S_{i}} \frac{C_i N_i}{\prod_a  D_{i,a}}\ ,
\end{equation}
with the meaning of each term explained as follows. The first summation of $\sigma_2$ runs over the permutations of the external legs. The summation over $\Gamma_i$ means to sum over all
possible trivalent graphs, and the symmetry factors $S_i$ will remove the overcounting from the automorphism symmetry of the graphs. The $C_i$ and $N_i$ correspond to the color factor and kinematic numerator of the $i$th trivalent graph, and $1/D_{i,a}$ denotes the $a$th propagator of the $i$th graph. 

\begin{figure}[t]
	\centerline{\includegraphics[height=2.8cm]{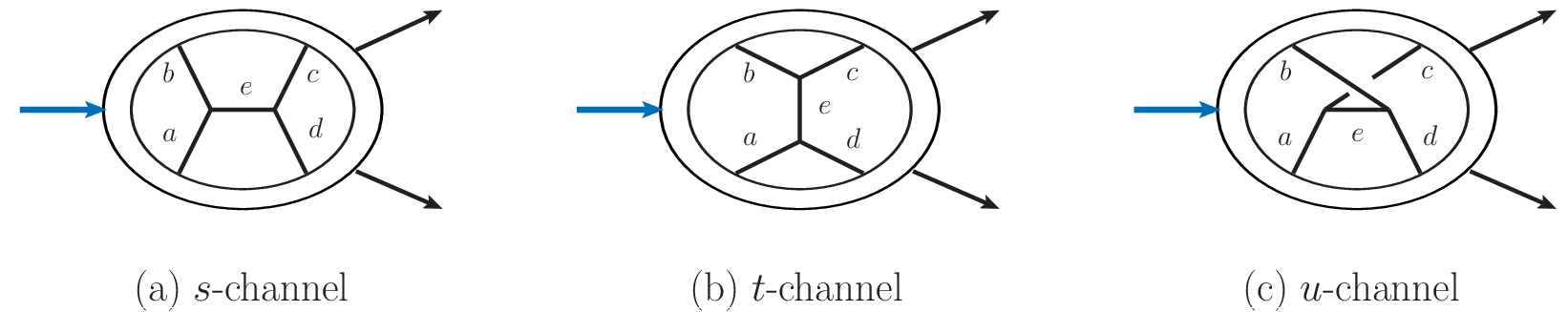} } 
	\caption{CK relation at loop-level for form factor.} 
	\label{FF_loop_diagrams_related_by_jacob}
\end{figure}

To impose CK duality at the loop level, one can pick one propagator and select the corresponding three $s, t, u$-channel topologies as in Figure~\ref{FF_loop_diagrams_related_by_jacob}. For these three graphs, their color factors have the form:
\begin{equation}
	C_s= f^{abs}f^{scd}(\delta \prod{f}), \quad C_t= f^{bct}f^{tda}(\delta \prod{f}), \quad C_u= f^{acu}f^{ubd}(\delta \prod{f}),
\end{equation}
where $\delta$ represents the color factor of the operator. The CK duality imposes that 
\begin{equation}\label{loop_stu_Jacobi}
	C_s= C_t + C_u  \quad \Rightarrow \quad N_s= N_t + N_u \,.
\end{equation}

The integrand that obeys all these identities will be referred to as ``global CK integrand''. In practical, we can construct the ``global CK ansatz'' for $N_i$ by following steps \cite{Bern:2012uf, Boels:2012ew, Carrasco:2015iwa}:
\begin{enumerate}
	\item 
	Generate all relevant trivalent diagrams.  
	\item 
	Generate all CK-dual Jacobi relations and find ``master topologies''. The master topologies represent a minimal set of topologies that can generate all other diagrams through dual Jacobi relations. The choice of master topologies is generally not unique.
	\item 
	Construct an ansatz for the numerators of master topologies. The ansatz in general depends linearly on a set of free parameters. Once we have the ansatz for master topologies, we can obtain all other numerators by CK relations. 
	\item 
	Apply various constraints, such as the symmetry properties (requiring each numerator to reflect the symmetry of the topology) and unitarity-cut constraints (see below), to solve for the ansatz. 
\end{enumerate}
The strategy of applying CK duality, if applicable, has many benefits. First, it will reduce the number of free parameters in the ansatz significantly, since we only made ansatz for ``master topologies''. Second, by construction, it captures both planar and non-planar structures.
Moreover, for amplitudes, we can directly perform double copy and get the corresponding gravity amplitude.

However, the main challenge is : it is not guaranteed that the ansatz will satisfy all the unitarity-cut constraints. In such a situation, one may try to modify or enlarge the ansatz and try again. Alternatively, one can introduce deformation for the failed cuts which we will use in the paper. We briefly review the unitarity method and deformation strategy in the following two subsections.

\subsection{Generalized unitarity method} 

To construct perturbative amplitudes or form factors, a powerful modern framework we can use is the unitarity-cut method \cite{Bern:1994ucl, Bern:1994cg, Britto:2005gu}. The key idea is to perform \textsl{cut} on an internal propagator, which means setting the propagator to be on-shell:
\begin{equation}
	\frac{i}{l^2} \stackrel{\text{cut}}{\longrightarrow} 2 \pi \delta_+(l^2)\ .
	\label{eq:cut}
\end{equation}
Under multiple cuts, a loop-level amplitude or form factor will factor into the product of tree-level building blocks, such as
\begin{equation}
	\itbf{F}_2^{(l)}|_{\text{cuts}} = \sum_{\text{physical states}} \prod (\text{tree-level blocks})\ ,
	\label{eq:Flcut}
\end{equation}
where the summation of internal physical states is needed. 
In $d$-dimension pure YM theory, this corresponds to contracting the polarization vectors of the cut internal propagator $l$ as 
\begin{equation}
	\sum_{\text{physical states}} \varepsilon^\mu(l) \varepsilon^\nu(l) = \eta^{\mu \nu} - \frac{l^\mu \xi^\nu + l^\nu \xi^\mu}{l \cdot \xi}\ ,
	\label{eq:helicities-sum}
\end{equation}
where $\xi^\mu$ is a light-like reference momentum.
A physical form factor is confirmed by verifying its consistency in all possible cut channels. Therefore, an efficient way to build loop-level form factors is by directly comparing the ansatz mentioned in the last subsection to results built by tree-level blocks. 

For the sake of convenience, one can consider color-ordered cuts with color-stripped tree-level blocks. Hence, we can perform color decomposition on both sides of (\ref{eq:Flcut}) before we actually use it to apply unitarity constraints. For instance, the LHS 
full-color form factor integrand can be decomposed into 
\begin{equation}
	\sum_{\Gamma_j} \left. \frac{C_j N_j}{\prod_a  D_{j,a}} \right|_{\text{cut}} = \sum_{\Gamma_j} \sum_{\sigma_1 \sigma_2 \dots \sigma_{n_i}} \text{tr}(\sigma_1)\text{tr}(\sigma_2) \dots \text{tr}(\sigma_{n_i}) \left. \frac{ N_j}{\prod_a  D_{j,a}} \right|_{\text{cut}}\ ,
\end{equation}
while the RHS reduces to
\begin{equation}
	\sum_{\sigma_1 \sigma_2 \dots \sigma_{n_i}} \text{tr}(\sigma_1)\text{tr}(\sigma_2) \dots \text{tr}(\sigma_{n_i}) F^{(0)}(\sigma_1) A^{(0)}(\sigma_2) \dots A^{(0)}(\sigma_{n_i})\ ,
\end{equation}
where both $F^{(0)}$ and $A^{(0)}$ are color-ordered tree blocks. Now unitarity constraints can be extracted by matching the coefficient of each multi-trace base. 
We mention that recently some progress in evaluating the full-color unitarity cuts was made in \cite{Carrasco:2024knk, Bern:2024vqs}.

\subsection{Deformation method}
Despite many favorable properties of imposing CK duality mentioned earlier, the duality is still a conjecture at the loop level. In particular, it can be hard to construct CK integrand for high-loop or high-point cases by simply enlarging the minimal ansatz.
A new strategy of applying minimal deformation was introduced in \cite{Li:2023akg}, which was found to be very effective for the two-loop four-point amplitude in pure YM theory, and a very compact solution was obtained.
Below we briefly review the core idea of this deformation method, which will be applied later to the three-loop Sudakov form factor.

The deformation method has two main steps: (1) construct the global CK-dual numerators and identify the contradiction with the unitarity-cut constraints, and (2) introduce deformation satisfying a subset of dual Jacobi relation and solve for the deformation to get the physical solution. 
Let us explain this in more detail.

The first step is constructing the global CK numerators $n_i$ following the standard steps reviewed above. When a physical solution can not be obtained, one can identify the ``failed" cuts that $n_i$ can not satisfy directly. 
We denote the set of such cuts as $\mathcal{U}_1$. The other cuts that can be satisfied directly are denoted as $\mathcal{U}_2$. The topologies that will contribute to cuts in $\mathcal{U}_1$ are the objects that are to be deformed, and we collect these topologies as set $\mathcal{T}$.

Second, we introduce the deformed numerators $N_i$ as
\begin{equation}
	N_i = \left\{ \begin{matrix} & n_i +\Delta_i, & \qquad\qquad i \in \mathcal{T} \\ & n_i , & \qquad\qquad i \notin \mathcal{T} \end{matrix} \right.
\end{equation} 
where deformation $\Delta_i$ is introduced to numerators in $\mathcal{T}$ and numerators outside $\mathcal{T}$ remain unchanged. To maximize the use of CK duality, we impose off-shell CK relations among $\Delta_i$. 
The dual Jacobi relations imposed on $\Delta_i$ are restricted in $\mathcal{T}$. CK relations that involve topologies that do not belong to $\mathcal{T}$ will be excluded. Practically, this can be achieved by ruling out the CK propagators which are cut by $\mathcal{U}_1$.
Using these relations we can again choose a minimal set of ``master topologies'' for the deformation.

Next, we make ansatz for the master deformed topologies. Note that the goal is to have a physical solution consistent with all cuts, so one can always make simple assumptions to simplify the ansatz. For example, we can ask that all $\Delta_i$ should not affect the cuts in $\mathcal{U}_2$ that are already satisfied by $n_i$. One may also ask the deformation for non-planar topologies to be zero and check planar-ordered cuts first.
Finally, one can solve the deformation ansatz by requiring $N_i$ to satisfy the previously failed unitarity cuts. 

In the next two sections, we will see that the above procedure works very well for the three-loop Sudakov form factor.

\section{Global CK-dual construction}
\label{sec:ff3loop}

Following the strategy explained in the last section, we consider a global CK-dual minimal ansatz for the three-loop Sudakov form factor and we will identify the inconsistency with unitarity cuts.

\subsection{CK-dual ansatz}

\begin{figure}[htbp]
    \centering
    \begin{minipage}{1.\textwidth}
        \centering
        \includegraphics[width=\textwidth]{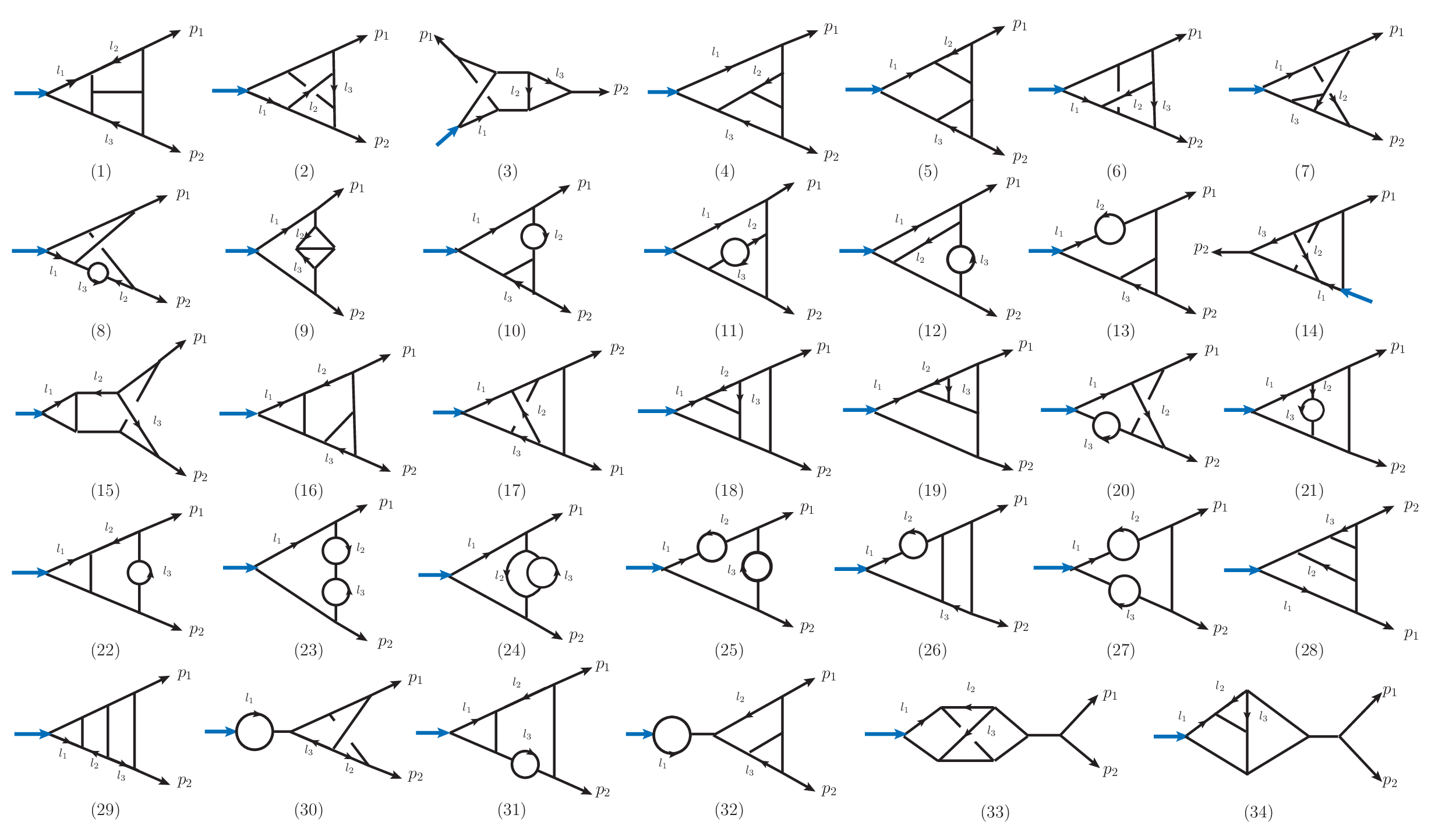}
    \end{minipage}
    \vspace{-0.2cm} 

    \begin{minipage}{1.\textwidth}
        \centering
        \includegraphics[width=\textwidth]{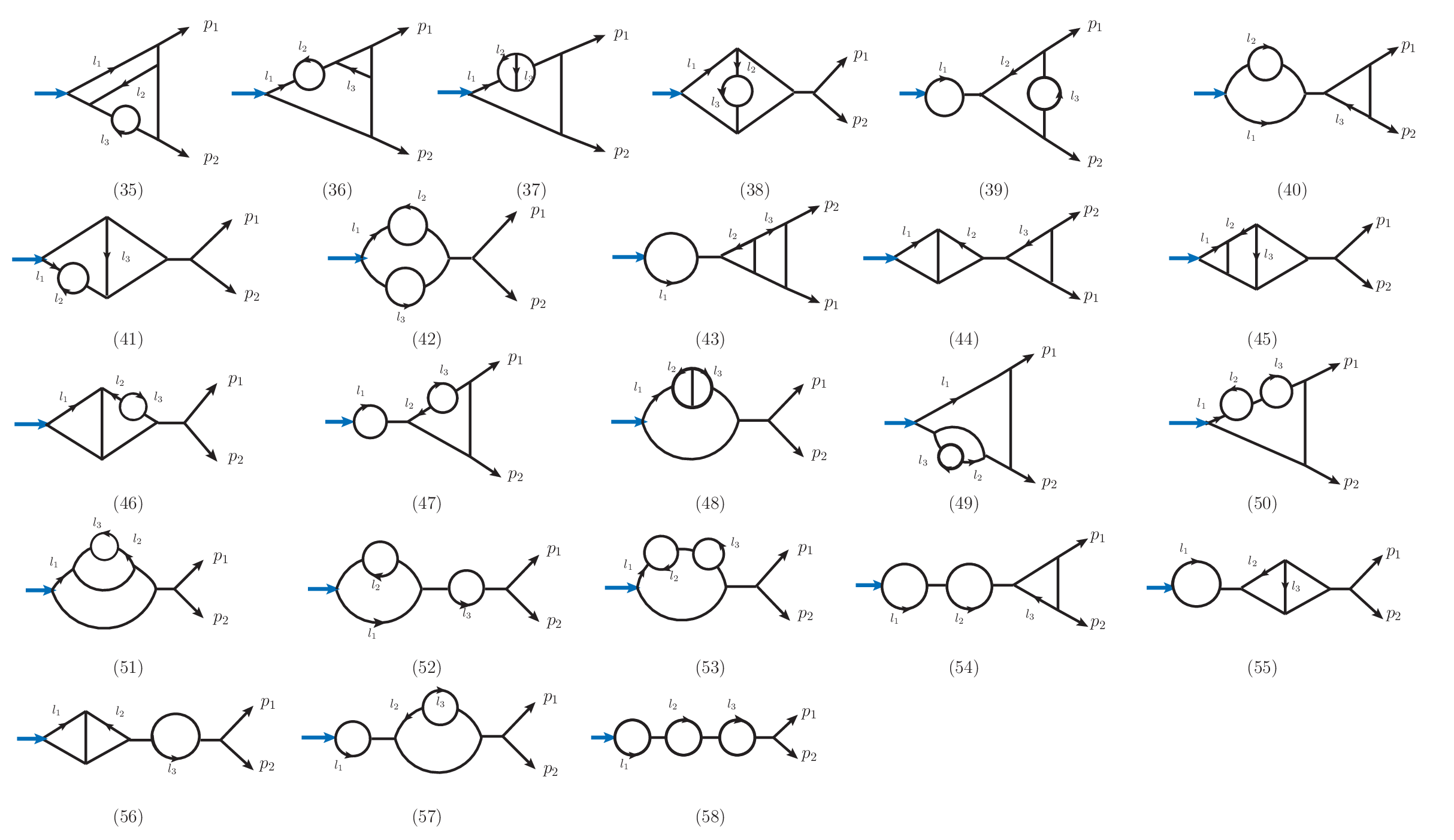} 
    \end{minipage}
    \caption{Trivalent topologies for the three-loop Sudakov form factor.}
    \label{fig:3_loop_all_topologies}
\end{figure}

We first collect all the trivalent topologies that will be considered in the following process in Figure~\ref{fig:3_loop_all_topologies}. 
We comment that trivalent topologies including tadpole or massless-bubble sub-graphs are ignored since they will vanish after integration, and CK relations related to these topologies will not be taken into account.

\begin{figure}[t]
	\centerline{\includegraphics[height=2.7cm]{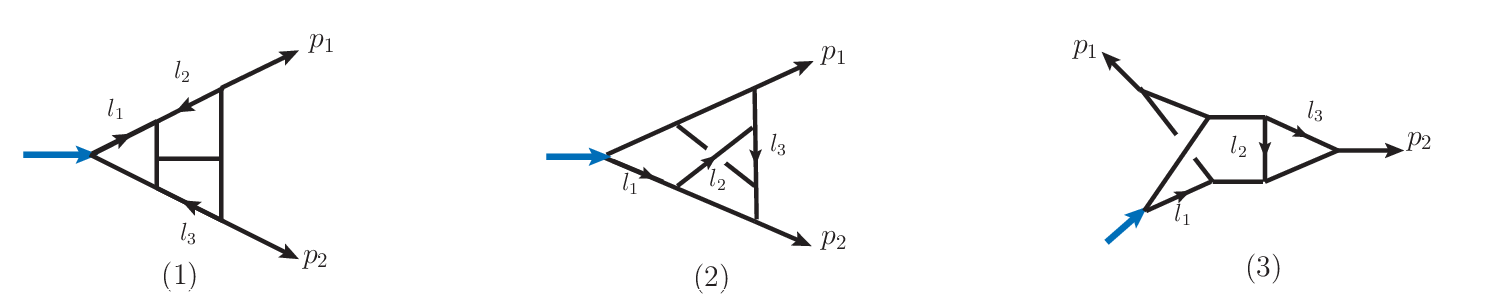} } 
	\caption{The master topologies for three-loop sudakov form factor.} 
	\label{fig:3_loop_master_topologies}
\end{figure}

Second, a chain of dual Jacobi relations is generated, which enables the deduction of all other numerators from three master topologies.  Here, the topologies (1), (2), and (3), collected also in Figure~\ref{fig:3_loop_master_topologies}, are chosen to be the master topologies.
The explicit set of dual Jacobi equations is given in Appendix~\ref{app:three_loop_CKrelation}.

Next, we make ansatz for the three master numerators, denoted as $n_1$, $n_2$ and $n_3$, 
\begin{equation}
	\label{formfactor_ansatz}
	n_{m} =  \sum_{k} a_{mk} M_{k} \,, \qquad m=1,2,3 \,
\end{equation}
where $a_{mk}$ are the coefficient to be determined, and $M_{k}$ are monomials consisted of the Lorentz products 
\begin{equation}\label{ff_lorntzproduct_basis}
	\{ \varepsilon_1 \cdot \varepsilon_2, \; \varepsilon_i \cdot p_j, \; \varepsilon_i \cdot l_\alpha, \; p_i \cdot l_\alpha, \; l_\alpha \cdot l_\beta, \; p_1 \cdot p_2  \} \,,
\end{equation}
with $i=1,2$ and $\alpha,\beta = 1,2,3$. Each $M_k$ has mass dimension 8. 
It is convenient to divide the terms in ansatz into two parts according to the structure of polarization vector $\varepsilon_1$ and $\varepsilon_2$:
\begin{equation}\label{eq:classification_of_FF_terms}
	n_m = n_m^{[1]}+n_m^{[2]} \;,
\end{equation}
Terms which are proportional to $\varepsilon_1 \cdot \varepsilon_2$ are collected in $n_m^{[1]}$ and all others belong to $n_m^{[2]}$:
\begin{equation*}
	\begin{aligned}
	& n^{[1]}: \quad (\varepsilon_1 \cdot \varepsilon_2) \times \textrm{(polynomials of momentum contractions)} , \\
	& n^{[2]}: \quad \textrm{Other terms}.
	\end{aligned}
\end{equation*}
Under $d$-dimensional unitarity cuts, they can be compared separately with their corresponding part of tree products. 
Note that the numerator ansatz is expressed in a polynomial local form using Lorentz products \eqref{ff_lorntzproduct_basis}. 
To provide more detail on the structure of the ansatz: each $n_m^{[1]}$ contains 1820 parameters, while each $n_m^{[2]}$ contains 7280 parameters. Consequently, there are 9100 parameters for each master topology, resulting in a total of 27300 parameters.\footnote{For convenience, here we only impose the dimensional power-counting constraint for the numerators in terms of the Lorentz products. The ansatz could be further refined by considering the power counting for loop momenta and excluding terms that reduce to tadpole or massless bubble contributions.}

Before considering unitarity cuts, we impose the symmetry constraints that the numerator should respect the automorphism symmetry of the corresponding graph. The number of parameters will reduce to 6662 after all the symmetry constraints are satisfied, 1449 for $n_m^{[1]}$ and 5213 for $n_m^{[2]}$. We checked that once all the symmetry constraints are satisfied, $n_i$ generated by the CK chain will automatically satisfy the full set of Jacobi relations.

\subsection{Unitarity constraint}
\begin{figure}[t]
	\centerline{\includegraphics[height=8cm]{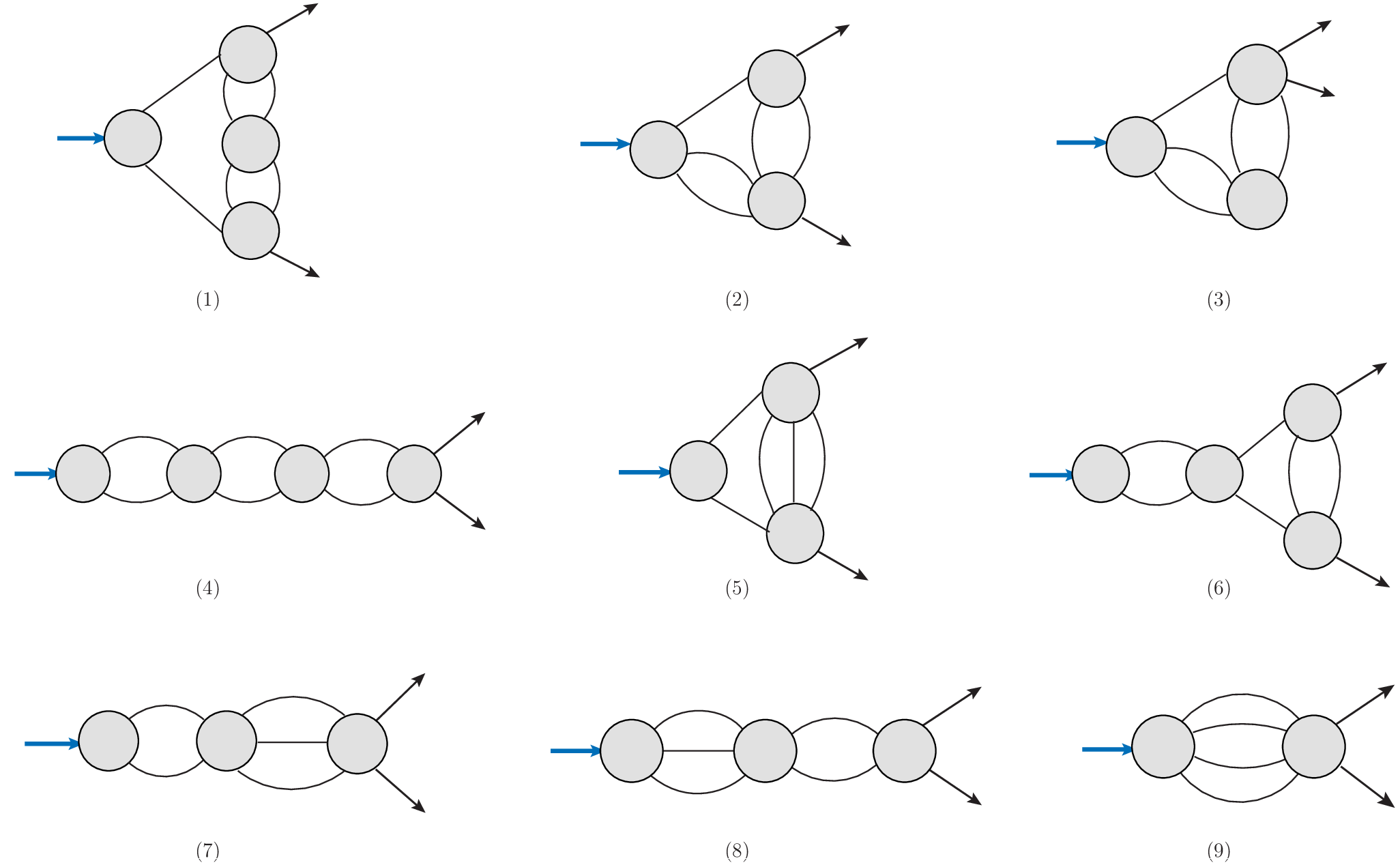} } 
	\caption{The spanning set of cuts for the three-loop Sudakov form factor.} 
	\label{fig:3_loop_spanning_cuts}
\end{figure}

The next step is to take unitarity cuts for the integrands and compare them with the tree products. The complete spanning set of cuts for the three-loop Sudakov form factor is shown in Figure~\ref{fig:3_loop_spanning_cuts}. These cuts are in general full-color cuts where each tree amplitude contains all possible orders.

\begin{figure}[t]
	\centerline{\includegraphics[width=0.45\linewidth]{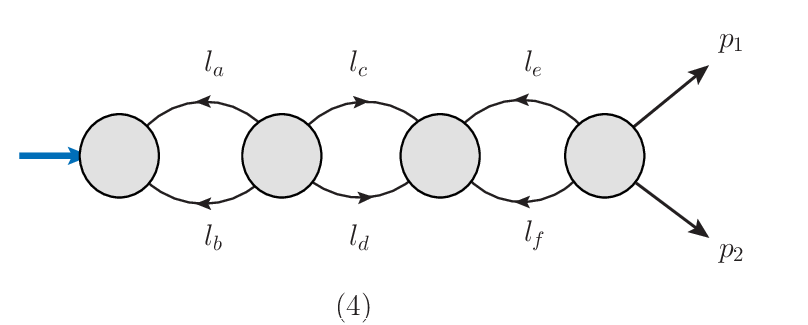} } 
	\caption{The planar order considered for cut (4).} 
	\label{fig:3_loop_cut_4_planar}
\end{figure}

\begin{figure}
	\begin{center}
		\includegraphics[width=0.95\textwidth]{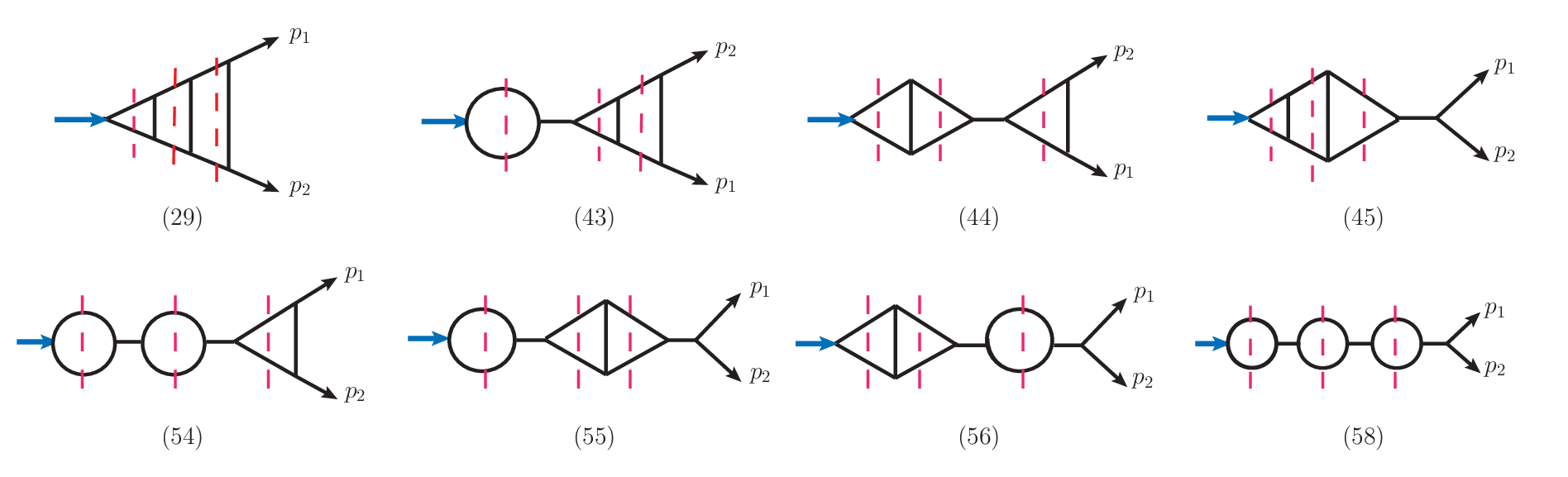}
	\end{center}
	\caption{Topologies detected by the special planar order of cut (4)}\label{fig:3_loop_cut_4_related}
\end{figure}

As a simple example, a planar color-ordering  cut (4) is shown in Figure~\ref{fig:3_loop_cut_4_planar} which is associated with the color factor 
\begin{equation}
\text{tr}(l_a, l_b) \text{tr}(l_c, l_d, l_b, l_a) \text{tr}(l_e, l_f, l_d, l_c) \text{tr}(1, 2, l_f, l_e) \,.
\end{equation} 
This cut has contributions from eight topologies shown in Figure~\ref{fig:3_loop_cut_4_related}. 
Thus, the unitarity constraint is built by matching summation over these topologies to the tree products
\begin{equation}
	F^{(3)}_2(1,2)|_{\text{cut-}(4)} = \sum_{{\rm physical \; states} } F_2^{(0)}(l_a,l_b) A_4^{(0)}(l_c, l_d, l_b, l_a) A_4^{(0)}(l_e, l_f, l_d, l_c) A_4^{(0)}(1, 2, l_f, l_e)\; .
\end{equation}

Note also that the coefficients $a_{mk}$ are in general polynomials of the space-time dimension parameter $d$. In the tree products, $d$ can be generated by the contraction of metric $\eta_{\mu \nu} \eta^{\mu \nu} = d$ during the summation of internal helicities in loop structures by using (\ref{eq:helicities-sum}). For the cut (4), terms like 
\begin{equation}
\eta_{l_a l_b}\, \eta^{l_a l_b}\, \eta_{l_c l_d}\, \eta^{l_c l_d}\, \eta_{l_e l_f}\, \eta^{l_e l_f} = d^3 \,,
\end{equation} 
will appear, together with lower-order terms. The general form for the coefficients of the three-loop Sudakov form factor is
\begin{equation}
	a_{mk} = \sum_{j=0}^{3} a_{mk, j} d^j,\ j = 0, 1, 2, 3 \; ,
	\label{eq:coefficients_general_form}
\end{equation}
where $a_{mk, j}$ are pure numbers. 

As for the two-loop four-gluon amplitude, we find that the global CK-dual integrand for the three-loop Sudakov form factor can not pass all the unitarity cuts. 
This is not beyond expectation since one can relate the three-loop Sudakov form factor to the four-point two-loop amplitude by cutting the two propagators connected with the operator. 
In other words, the four-point amplitude should appear as a substructure in the three-loop Sudakov form factor and the property arising in the four-gluon amplitude should also appear in the latter. 
A more detailed discussion along this connection will be given in Section~\ref{sec:discussion}.

However, there are also important differences. While the two-loop amplitude has only one failed planar-ladder two-double cut \cite{Bern:2015ooa, Li:2023akg, Edison:2023ulf}, we find that the three-loop form factor has three types of unitarity cuts failed, the cut (1), (2) and (3) in Figure~\ref{fig:3_loop_spanning_cuts}. 
More specifically, $n_i^{[1]}$ fails to pass the cut (1), and $n_i^{[2]}$ can not pass the cut (1), (2) and (3). 
Therefore, the three-loop form factor is much more complicated and has certainly new structures beyond the two-loop case.
Our next goal is to apply the minimal deformation strategy to rectify these failed cuts.

\section{Deformation}\label{sec:FFdeformation}
In this section, we introduce deformations to $n_i$ and obtain new numerators $N_i$ that can pass all unitarity cuts and also satisfy on-shell CK relations. 
Concretely, we define
\begin{equation}\label{eq:3_loop_deform_integrand}
	N_i = \left\{ \begin{matrix} & n_i +\Delta_i, & \qquad\qquad i \in \{\textrm{cut(1),(2),(3)-related topologies}\},  \\ & n_i , & \textrm{other topologies}, \end{matrix} \right.
\end{equation} 
where $\Delta_{i}$ is the deformation to be solved. 
As for the decomposition of $n_i$ in \eqref{eq:classification_of_FF_terms}, we also divide $\Delta_{i}$ into $\Delta_i^{[1]}$ and $\Delta_i^{[2]}$ according to the Lorentz structure for the polarization vectors as
\begin{align}
	& \Delta^{[1]}: \quad (\varepsilon_1 \cdot \varepsilon_2) \times \textrm{polynomials of momentum contractions} \,, \nonumber\\
	& \Delta^{[2]}: \quad \textrm{Other terms} \,. \nonumber
\end{align}

We would like to explain our strategy of choosing the ansatz for the deformation. We will assume that CK duality is only `weakly' violated, and in such a scenario, our goal is to identify the `simplest' deformation that can yield a consistent physical solution. In the following analysis, we will adhere first to the planar-type cuts and assume that the deformation related to non-planar topologies is negligible. Moreover, we will impose further constraints on the ansatz for the deformed numerators such that they do not affect the previously satisfied cuts. These considerations will significantly reduce the parameter space of the ansatz and can be taken as an assumption for the simplicity of the solution. Fortunately, we find that this assumption is always valid for the form factor we consider.

\subsection{$\Delta_i^{[1]}$ part}

In this subsection, we consider first the simpler $\Delta_i^{[1]}$ part.
From the last section, we know that for the global ansatz $n_i^{[1]}$, only the cut (1) in Figure~\ref{fig:3_loop_spanning_cuts} can not pass.
So we need to introduce deformation to amend this cut. 

\begin{figure}[t]
	\centerline{\includegraphics[height=5.8cm]{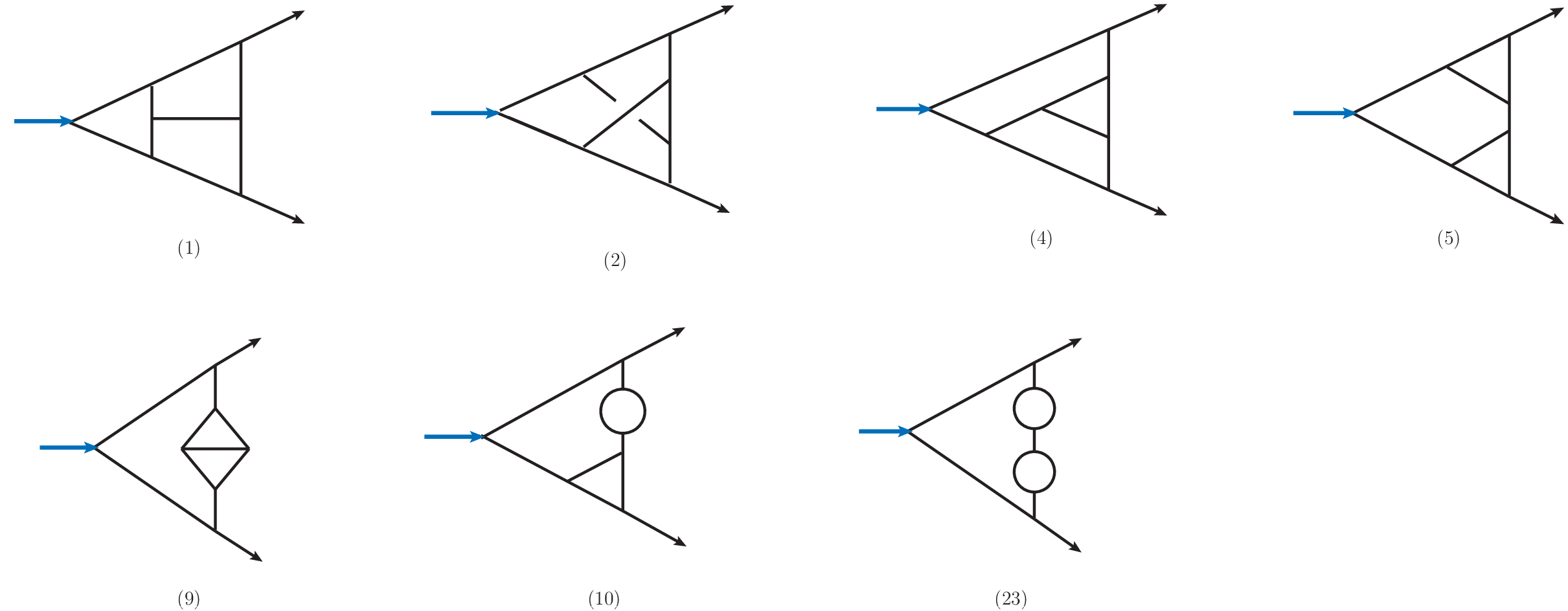} } 
	\caption{Topologies that can be detected by  cut (1) in Figure~\ref{fig:3_loop_spanning_cuts}.} 
	\label{fig:3_loop_deformed_topologies_1}
\end{figure}

The first step is to identify the topologies related to this cut.
In Figure~\ref{fig:3_loop_deformed_topologies_1}, we collect the topologies that can be detected by the cut (1).

Among the topologies in Figure~\ref{fig:3_loop_deformed_topologies_1}, we find that the topology (1) and (2) can serve as master topologies, and other $\Delta_i^{[1]}$ can be obtained by following CK relations:
\begin{equation}\label{eq:CK_relations_for_deform1}
	\begin{aligned}
		\Delta_{4}^{[1]}&=\Delta_{2}^{[1]}[p_2,p_1,l_1,l_1-l_2-p_1,l_1 - l_2 - l_3 - p_1-p_2]-\Delta_{1}^{[1]}[p_1,p_2,l_1,l_2-l_1,l_3]\\
		\Delta_{5}^{[1]}&=\Delta_{1}^{[1]}-\Delta_{2}^{[1]}[p_1,p_2,p_1+p_2-l_1, p_1 + p_2 - l_1+l_3, p_2 -l_1 - l_2 + l_3 ]\\
		\Delta_{9}^{[1]}&=-\Delta_{4}^{[1]}[p_1,p_2,l_1, l_2, l_1 + l_3 - p_1 - p_2]-\Delta_{4}^{[1]}[p_1, p_2, l_1, l_1 - l_2 - p_1, -l_3 - p_2]\\
		\Delta_{10}^{[1]}&=-\Delta_{4}^{[1]}-\Delta_{4}^{[1]}[p_1, p_2, l_1, l_1 - l_2 - p_1, l_3]\\
		\Delta_{23}^{[1]}&=\Delta_{9}^{[1]}+\Delta_{9}^{[1]}[p_1, p_2, l_1, l_1 - l_2 - p_1, l_3]\\
	\end{aligned}
\end{equation}
We stress that no cut condition is imposed on these relations.
To select the above dual Jacobi relations, it is important to ensure that they do not extend to other topologies beyond Figure~\ref{fig:3_loop_deformed_topologies_1}.
Practically, this can done by avoiding applying Jacobi operation on propagators that are severed by cut (1).

Next, we make an ansatz for $\Delta_{1}^{[1]}$ and $\Delta_{2}^{[1]}$. 
As mentioned at the beginning of this section, we would like to find the deformation as simple as possible. We observe that the second master is of non-planar topology, and in addition, it does not contribute to the planar cut discussed below. Therefore, we set $\Delta_{2}^{[1]}$ to 0 and only need to make ansatz for $\Delta_{1}^{[1]}$.

\begin{figure}[t]
	\centerline{\includegraphics[height=3.4cm]{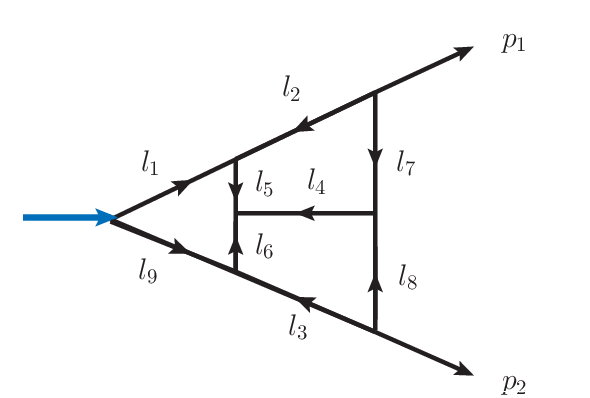} } 
	\caption{The master deformed topology and the momentum labeling.} 
	\label{fig:3_loop_tp1_label_all}
\end{figure}

To avoid affecting the already satisfied cuts, we assume that $\Delta_{1}^{[1]}$ should directly vanish in those cuts. We find that a simple way to achieve this is to require $\Delta_{1}^{[1]}$ to be proportional to $l_2^2$, $l_3^2$ and $l_4^2$ at the same time, with momentum labeling shown in Figure~\ref{fig:3_loop_tp1_label_all}. So we propose the ansatz of $\Delta_{1}^{[1]}$ as
\begin{equation}\label{eq:3_loop_deform_ansatz_1}
	\Delta_{1}^{[1]} =  (\sum_{k} c_{k}^{[1]} M_{k}^{[1]}) (\varepsilon_1 \cdot \varepsilon_2) \, l_2^2 l_3^2 l_4^2  \;,
\end{equation}
where $M_{k}^{[1]}$ are monomials formed by the product of following basis
\begin{equation}
	\{p_i \cdot l_\alpha, \; l_\alpha \cdot l_\beta, \; p_1 \cdot p_2  \}, \;
\end{equation}
with $i=1,2$ and $\alpha,\beta = 1,2,3$. A simple dimensional analysis shows that $M_{k}^{[1]}$ has mass dimension 2, and there are 13 parameters $c_{k}^{[1]}$ in total. After imposing the symmetry property for the numerator $\Delta_{1}^{[1]}$, only 8 parameters remain. Given this ansatz, we obtain the deformation $\Delta_i^{[1]}$ of other topologies, and we verify that they directly vanish under all other cuts and satisfy the corresponding symmetry property.

\begin{figure}[t]
		\centerline{\includegraphics[height=4.5cm]{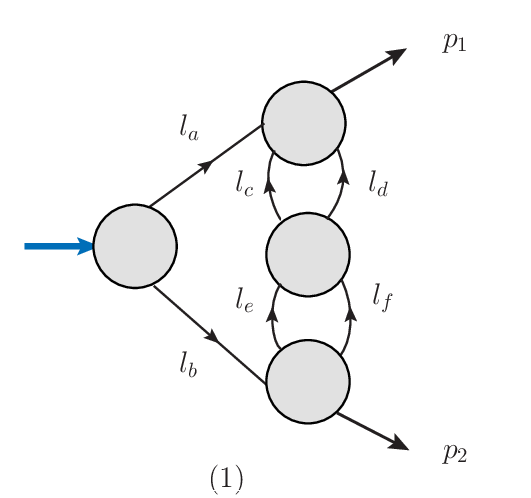} } 
		\caption{The planar cut for the cut-(1) in  Figure~\ref{fig:3_loop_spanning_cuts}.} 
		\label{fig:cut_1_order}
\end{figure}

Now we would like to recheck the deformed ansatz for the planar cut (1) shown in Figure~\ref{fig:cut_1_order}:
\begin{equation}
		F^{(3)}_2(1,2)|_{\textrm{cut-}(1)} = \sum_{{\rm physical \; states} } F_2^{(0)}(l_a,l_b) A_4^{(0)}(2,-l_b,l_e,l_f) A_4^{(0)}(l_c,l_d,-l_f,-l_e) A_4^{(0)}(1,-l_d,-l_c, -l_a),
\end{equation}
which is associated with the color ordering ${\rm tr}(l_a,l_b) {\rm tr}(2,l_b,l_e,l_f){\rm tr}(l_c,l_d,l_f, l_e) {\rm tr}(1, l_d, l_c, l_a)$. 
It is easy to see that $\Delta_{2}^{[1]}$ for the topology (2) of Figure~\ref{fig:3_loop_deformed_topologies_1} does not contribute to this cut. 
To compare with the cut contribution from $N_i^{[1]}$, we extract the part that is proportional to $(\varepsilon_1 \cdot \varepsilon_2)$ in the tree product.

We find that the deformed ansatz can indeed satisfy this unitarity cut, which fixes one parameter. The solution automatically passes all other cuts. Thus, the final solution space of $\Delta_i^{[1]}$ contains 7 free parameters. 
Interestingly, within the solution space, an especially simple  solution of $\Delta_{1}^{[1]}$ can be chosen as:
\begin{equation}\label{eq:3_loop_deform_Delta_1}
	\Delta_{1}^{[1]} = 2(d-2)^2 (\varepsilon_1 \cdot \varepsilon_2)(p_1 \cdot p_2 ) l_2^2 l_3^2 l_4^2  \;.
\end{equation}

\subsection{$\Delta_i^{[2]}$ part}
\begin{figure}[t]
	\centerline{\includegraphics[height=9.9cm]{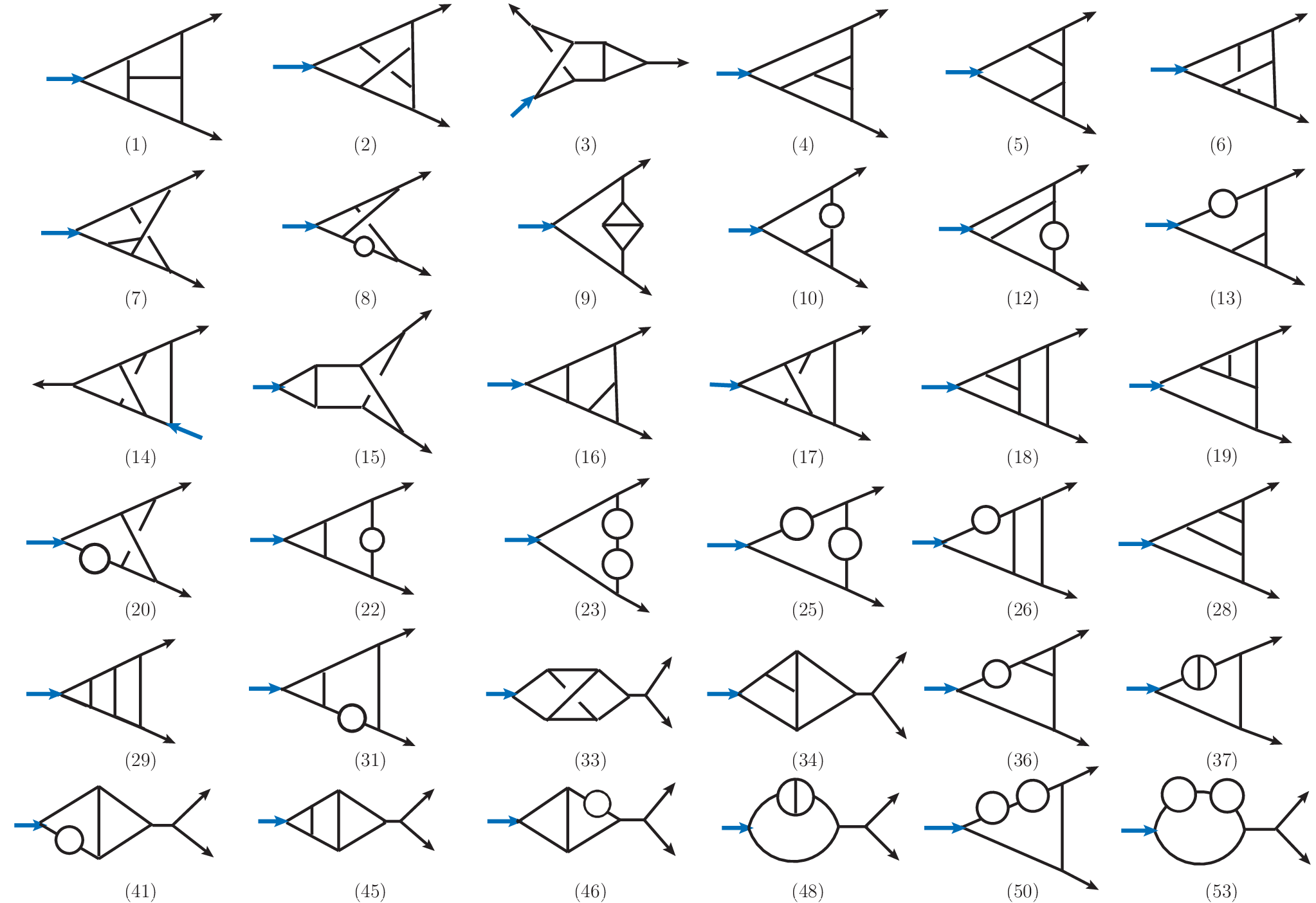} } 
	\caption{Topologies that can be detected  by the cuts (1), (2), or (3).} 
	\label{fig:3_loop_deformed_topologies_2}
\end{figure}

The second part $\Delta_{i}^{[2]}$ is more complicated since there are three failed cuts: the cuts (1), (2), (3) in Figure~\ref{fig:3_loop_spanning_cuts}. 
We need to make deformation for the numerators that can affect these cuts.
As before, we first collect all topologies that can be detected by the (full-color) cuts (1), (2), and (3) in Figure~\ref{fig:3_loop_deformed_topologies_2}. 

A set of CK chains can be generated that relate all of them. We find that we need at least three numerators to deduce all other $\Delta_i^{[2]}$. We choose $\Delta_{1}^{[2]}$, $\Delta_{2}^{[2]}$, and $\Delta_3^{[2]}$ as master, and the other $\Delta_i^{[2]}$ can be obtained via following relations:
\begin{align}
	\dt{4} &= -\dt{1}[p_1, p_2, l_1, -l_1 + l_2, l_3] + \dt{2}[p_2, p_1, l_1, l_1 - l_2 - p_1, l_1 - l_2 - l_3 - p_1 - p_2] \notag \\ 
	\dt{5} &= \dt{1} - \dt{2}[p_1, p_2, -l_1 + p_1 + p_2, -l_1 + l_3 + p_1 + p_2, -l_1 - l_2 + l_3 + p_2] \notag \\ 
	\dt{6} &= \dt{2}[p_2, p_1, -l_1 + p_1 + p_2, -l_1 - l_2 - l_3 + p_2, -l_1 - l_2] + \dt{3}[p_1, p_2, l_1, -l_1 - l_2, -l_3 + p_2] \notag \\ 
	\dt{7} &= \dt{2}[p_2, p_1, -l_1 + p_1 + p_2, -l_2 - l_3 + p_2, -l_3] - \dt{6}[p_2, p_1, -l_1 + p_1 + p_2, l_2 + l_3 - p_2, -l_3] \notag \\ 
	\dt{8} &= -\dt{3}[p_1, p_2, l_1, -l_3, l_2 + l_3 + p_2] - \dt{3}[p_1, p_2, -l_1 + p_1 + p_2, -l_2 - l_3, l_3] \notag \\ 
	\dt{9} &= -\dt{4}[p_1, p_2, l_1, l_2, l_1 + l_3 - p_1 - p_2] - \dt{4}[p_1, p_2, l_1, l_1 - l_2 - p_1, -l_3 - p_2] \notag \\ 
	\dt{10} &= -\dt{4} - \dt{4}[p_1, p_2, l_1, l_1 - l_2 - p_1, l_3] \notag \\ 
	\dt{12} &= \dt{4}[p_1, p_2, l_1, l_2 - l_3, l_1 - l_2 - p_1 - p_2] + \dt{4}[p_1, p_2, l_1, l_1 + l_3 - p_1, l_1 - l_2 - p_1 - p_2] \notag \\ 
	\dt{13} &= \dt{5}[p_2, p_1, -l_1 + p_1 + p_2, l_3, -l_1 - l_2] + \dt{5} \notag \\ 
	\dt{14} &= -\dt{6}[p_1, p_2, l_1, -l_1 - l_2, -l_3 + p_2] + \dt{6} \notag \\
	\dt{15} &= -\dt{6} - \dt{6}[p_2, p_1, l_1, l_2, -l_2 - l_3] \notag \\
	\dt{16} &= \dt{1} - \dt{6}[p_1, p_2, -l_1 + p_1 + p_2, -l_2 - l_3 - p_1 - p_2, l_3 + p_2] \notag \\ 
	\dt{17} &= \dt{6}[p_1, p_2, l_1, l_2, -l_2 - l_3 - p_1] + \dt{6}[p_1, p_2, -l_1 + p_1 + p_2, l_3, -l_2 - l_3 - p_1] \notag \\ 
	\dt{18} &= -\dt{1}[p_1, p_2, l_1, l_2, -l_2 - l_3 - p_1 - p_2] + \dt{6}[p_2, p_1, l_1, l_2, l_3 + p_1] \notag \\ 
	\dt{19} &= -\dt{7}[p_2, p_1, -l_1 + p_1 + p_2, l_2 + l_3 + p_1, -l_3] - \dt{18} \notag \\ 
	\dt{20} &= -\dt{7}[p_1, p_2, l_1, l_2, l_1 - l_2 - l_3 - p_1] - \dt{7}[p_2, p_1, l_1, l_1 - l_2, l_2 - l_3 - p_2] \notag \\
	\dt{22} &= \dt{16}[p_1, p_2, l_1, l_2, -l_3 - p_2] + \dt{16}[p_1, p_2, l_1, l_2, -l_2 + l_3 - p_1 - p_2] \notag \\ 
	\dt{23} &= \dt{9} + \dt{9}[p_1, p_2, l_1, l_1 - l_2 - p_1, l_3] \notag \\ 
	\dt{25} &= \dt{10}[p_2, p_1, -l_1 + p_1 + p_2, l_3, -l_1 - l_2] + \dt{10}[p_2, p_1, -l_1 + p_1 + p_2, l_3, l_2] \notag \\ 
	\dt{26} &= -\dt{18}[p_1, p_2, l_1, -l_1 - l_2, l_1 + l_2 - l_3 - p_1 - p_2] - \dt{18}[p_1, p_2, l_1, l_2, -l_2 - l_3 - p_1 - p_2] \notag \\ 
	\dt{28} &= -\dt{3}[p_1, p_2, l_1, -l_1 + l_2 + l_3 + p_1 + p_2, -l_3] - \dt{16}[p_1, p_2, l_1, -l_1 + l_2, l_3] \notag \\ 
	\dt{29} &= -\dt{17}[p_1, p_2, l_1,  l_2 + l_3, -l_2-p_1-p_2] - \dt{18}[p_1, p_2, p_1+p_2-l_1, -p_1-p_2-l_2, l_2+l_3] \notag \\ 
	\dt{31} &= \dt{16} + \dt{16}[p_1, p_2, l_1, l_2, -l_2 - l_3 - p_1 - p_2] \notag \\ 
	\dt{33} &= \dt{17} - \dt{17}[p_2, p_1, l_1, l_2, l_3] \notag \\ 
	\dt{34} &= \dt{18} - \dt{18}[p_2, p_1, l_1, l_2, l_3] \notag \\ 
	\dt{36} &= -\dt{19}[p_1, p_2, l_1, -l_1 - l_2, l_2 - l_3] - \dt{19}[p_1, p_2, l_1, l_2, -l_1 - l_2 - l_3] \notag \\ 
	\dt{37} &= -\dt{19}[p_1, p_2, l_1, -l_1 - l_2, -l_3] - \dt{19} \notag \\ 
	\dt{41} &= -\dt{26}[p_1, p_2, l_1, l_2, l_1 + l_3 - p_1 - p_2] + \dt{26}[p_2, p_1, l_1, l_2, l_1 + l_3 - p_1 - p_2] \notag \\ 
	\dt{45} &= \dt{29}[p_1, p_2, l_1, l_2, -l_2 - l_3] - \dt{29}[p_2, p_1, l_1, l_2, -l_2 - l_3] \notag \\ 
	\dt{46} &= \dt{45}[p_1, p_2, l_1, l_2, -l_2 - l_3] - \dt{45}[p_1, p_2, -l_1 + p_1 + p_2, -l_2 - p_1 - p_2, -l_3] \notag \\ 
	\dt{48} &= -\dt{37}[p_1, p_2, l_1, l_2, -l_2 - l_3] + \dt{37}[p_2, p_1, l_1, l_2, -l_2 - l_3] \notag \\
	\dt{50} &= \dt{36}[p_1, p_2, l_1, l_2, -l_3] + \dt{36}[p_1, p_2, l_1, l_2, -l_1 + l_3] \notag \\ 
	\dt{53} &= -\dt{50}[p_1, p_2, l_1, l_2, -l_3] + \dt{50}[p_2, p_1, l_1, l_2, -l_3] \,. \label{eq:CK_relations_for_deform2}
\end{align}
As before, no cut condition is imposed on these relations, and they do not involve topologies beyond those in Figure~\ref{fig:3_loop_deformed_topologies_2}.

\begin{figure}[t]
	\centerline{\includegraphics[height=2cm]{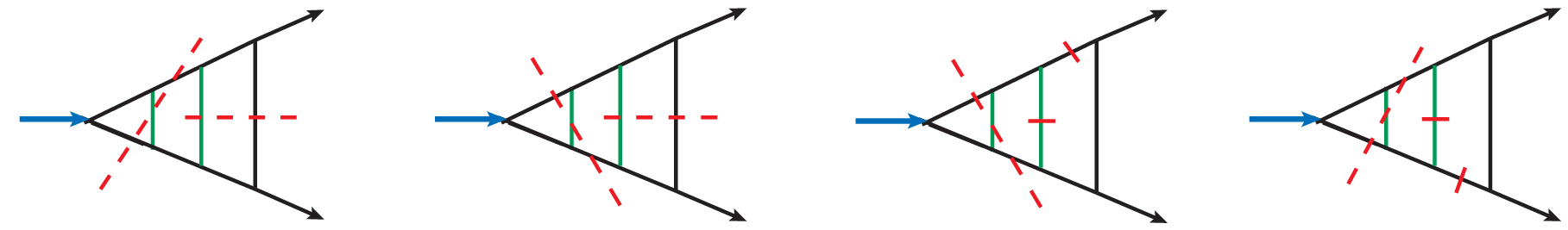} } 
	\caption{Topologies (29) possess 4 possible cut contributions to cut (2) and (3). We can see that the two green-line propagators are cut in all cases, thus Jacobi relations with respect to these two propagators should be excluded for $\Delta_i^{[2]}$.} 
	\label{fig:simu_cut_propagator}
\end{figure}

To select the proper Jacobi relations for the deformation, one complication compared to the previous case is that now there are three failed cuts. In choosing the dual Jacobi relations of \eqref{eq:CK_relations_for_deform2}, one can apply the Jacobi relation on propagators as long as they are not cut simultaneously by the three cuts.
As an example, in Figure \ref{fig:simu_cut_propagator}, we show the possible cuts for topology (29). We can see the two propagators indicated by the green color are severed by all four types of cuts. Therefore, we should not apply the Jacobi operation to these two propagators.

Among the three masters, $\Delta_{2}^{[2]}$ and $\Delta_{3}^{[2]}$ are of non-planar topology, and by the assumption of simplicity, we set them to be 0:
\begin{equation}
\Delta_2^{[2]} = 0 = \Delta_3^{[2]}  \,.
\end{equation}
With this assumption, we find that all other non-planar topologies in Figure~\ref{fig:3_loop_deformed_topologies_2} have zero deformation using dual Jacobi relations \eqref{eq:CK_relations_for_deform2}.

Now we have only one non-zero master numerator $\Delta_{1}^{[2]}$ for which we make an ansatz.
Similar to the case of $\Delta_1^{[1]}$, we assume the master deformation $\Delta_{1}^{[2]}$ vanishes directly in all other already satisfied cut. An analysis of the topology structure shows that it can be achieved by requiring each term either be proportional to $l_2^2 l_4^2$ or $l_3^2 l_4^2$, where the momentum labeling is the same as in Figure~\ref{fig:3_loop_tp1_label_all}.
The ansatz is thus proposed as
\begin{equation}\label{eq:3_loop_deform_ansatz_2}
	\Delta_{1}^{[2]} = l_4^2 \Big[ \Big(\sum_{k} c_{k,1}^{[2]} M^{[2]}_{k} \Big) l_2^2 + \Big( \sum_{k} c_{k,2}^{[2]} M^{[2]}_{k} \Big) l_3^2 \Big] \,,
\end{equation}
where $M^{[2]}_{k}$ are monomials given by products of
\begin{equation}\label{3_loop_deform_ansatz}
	\{\varepsilon_i \cdot p_j, \; \varepsilon_i \cdot l_{\alpha}, \; p_i \cdot p_j, \; p_i \cdot l_{\alpha}, \; l_{\alpha} \cdot l_{\beta} \} \, ,
\end{equation}
with $i,j = 1,2$ and $\alpha , \beta = 1,2,3$. Each $M^{[2]}_{k}$ is linear with $\varepsilon_i$ and has mass dimension 4. 
We find there are 200 independent monomials, and the full ansatz has 400 free parameters represented by $c_{k,1}^{[2]}$ and $c_{k,2}^{[2]}$. By requiring the ansatz to satisfy the symmetry property of the master topology, we can fix 198 parameters. All other non-zero $\Delta_{i}^{[2]}$ are determined by $\Delta_{1}^{[2]}$ using the CK relations given in \eqref{eq:CK_relations_for_deform2}.

Now we have the ansatz for the deformation numerators $\Delta_{i}^{[2]}$ that depend on 202 free parameters. Before checking against the three failed cuts,
we would like to comment on two further special choices.

\begin{figure}[t]
	\centerline{\includegraphics[height=5cm]{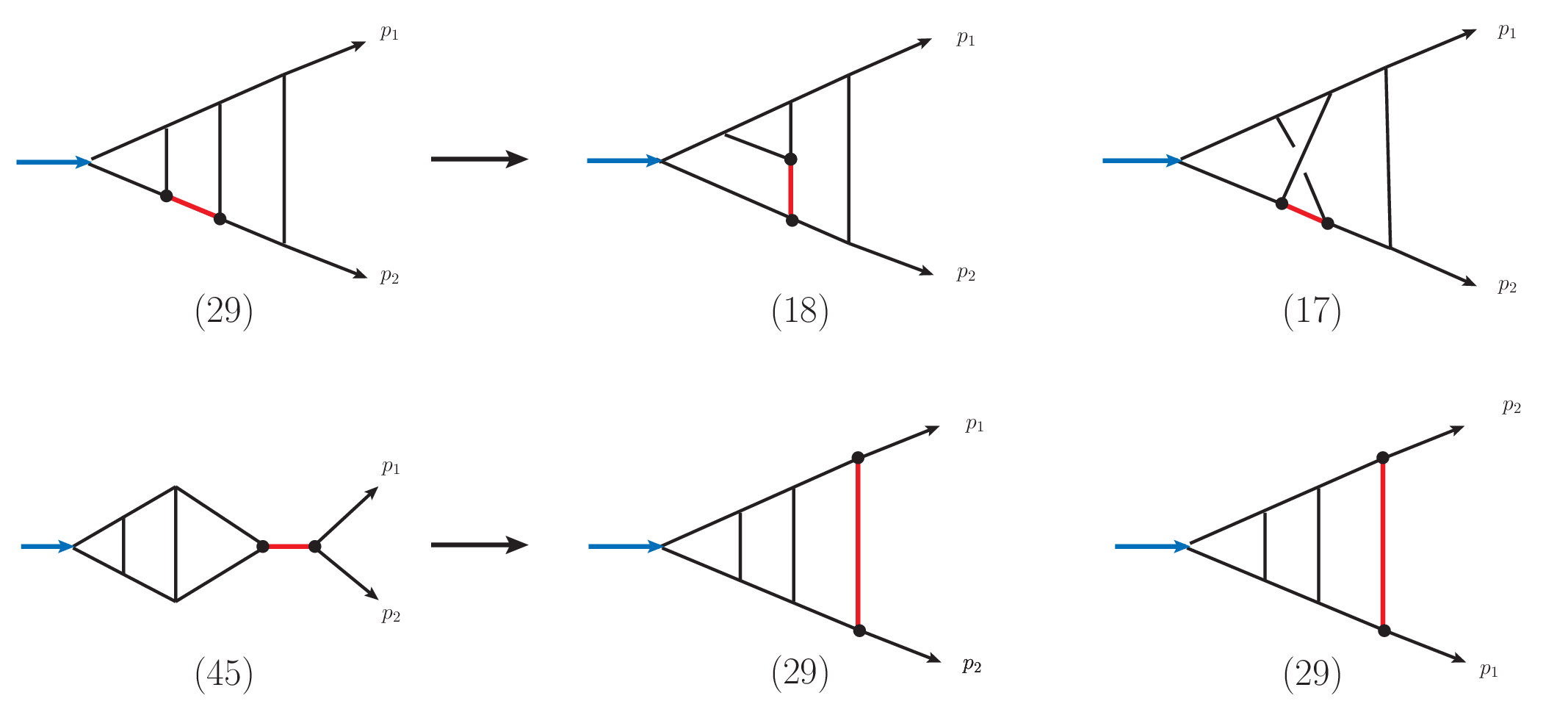} } 
	\caption{Dual Jacobi relations for $\Delta_{29}^{[2]}$ and $\Delta_{45}^{[2]}$. }
	\label{fig:3_loop_ck_example2}
\end{figure}

First, we check the other satisfied cuts, \emph{i.e.}, the cuts (4)--(9) in Figure~\ref{fig:3_loop_spanning_cuts}. 
We find that the deformed numerators all vanish on these cuts except $\Delta_{29}^{[2]}$ and $\Delta_{45}^{[2]}$, which have non-zero contributions to the cut (8) in Figure~\ref{fig:3_loop_spanning_cuts}. 
To understand this point, we recall that they are generated by the following dual-Jacobi relations
\begin{align}
& \dt{29} = -\dt{17}[p_1, p_2, l_1,  l_2 + l_3, -l_2-p_1-p_2] - \dt{18}[p_1, p_2, p_1+p_2-l_1, -p_1-p_2-l_2, l_2+l_3] , \label{eq:delta29Jacobi}\\
& \dt{45} = \dt{29}[p_1, p_2, l_1, l_2, -l_2 - l_3] - \dt{29}[p_2, p_1, l_1, l_2, -l_2 - l_3] \,, \label{eq:delta45Jacobi}
\end{align}
which are also shown in Figure~\ref{fig:3_loop_ck_example2}.
A natural step is to require them to vanish under the cut (8), and this will constrain the ansatz and reduce the number of free parameters to 87. However, we find the corresponding solution of $N_i^{[2]}$ can not pass other unitarity cuts. 
To resolve this problem, we choose ``not to impose" the off-shell dual Jacobi relation \eqref{eq:delta29Jacobi}. Moreover, as a choice of simplicity, we set $\Delta_{29}^{[2]}$ to 0. As we will discuss in Section~\ref{sec:onshellJacobi}, the Jacobi relation \eqref{eq:delta29Jacobi} will still be satisfied at the on-shell level. We assume the second Jacobi identity \eqref{eq:delta45Jacobi} still holds and $\Delta_{45}^{[2]}$ is also 0 correspondingly.

\begin{figure}[t]
	\centerline{\includegraphics[height=3cm]{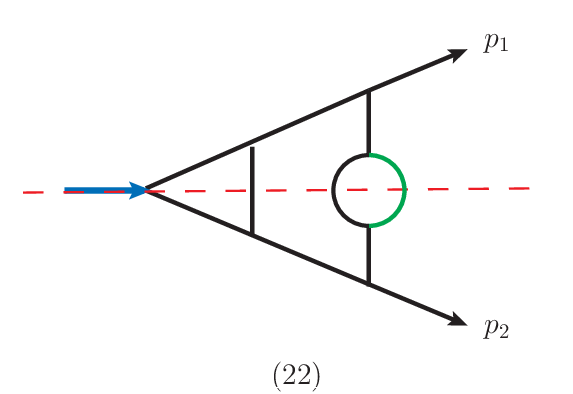} } 
	\caption{The horizontal flip symmetry property of topology (22) is not satisfied by $\Delta_{22}^{[2]}$. However, all terms in $\Delta_{22}^{[2]}$ are proportional to the green line propagator and reduce to a tadpole integral.} 
	\label{fig:3_loop_tp22_broken_sym}
\end{figure}

Second, we check the symmetry property for the $\Delta_{i}^{[2]}$. We find that, with $\Delta_{29}^{[2]}$ and $\Delta_{45}^{[2]}$ being 0, most of $\Delta_{i}^{[2]}$ automatically satisfy symmetry constraints except $\Delta_{22}^{[2]}$, generated by
\begin{equation}
\dt{22} = \dt{16}[p_1, p_2, l_1, l_2, -l_3 - p_2] + \dt{16}[p_1, p_2, l_1, l_2, -l_2 + l_3 - p_1 - p_2] \,. 
\end{equation}
This numerator breaks the horizontal flipping symmetry for the related graph as shown in Figure \ref{fig:3_loop_tp22_broken_sym}. 
It seems also natural to further impose this symmetry constraint to the ansatz, however, we find the resulting solution would be inconsistent with unitarity cuts. 
Interestingly, we observe that all terms in $\Delta_{22}^{[2]}$ can be reduced to scaleless integrals and consequently vanish in all unitarity cuts. 
Therefore, we will not impose this symmetry constraint in our calculation.

\begin{figure}[t]
	\centerline{\includegraphics[height=3.2cm]{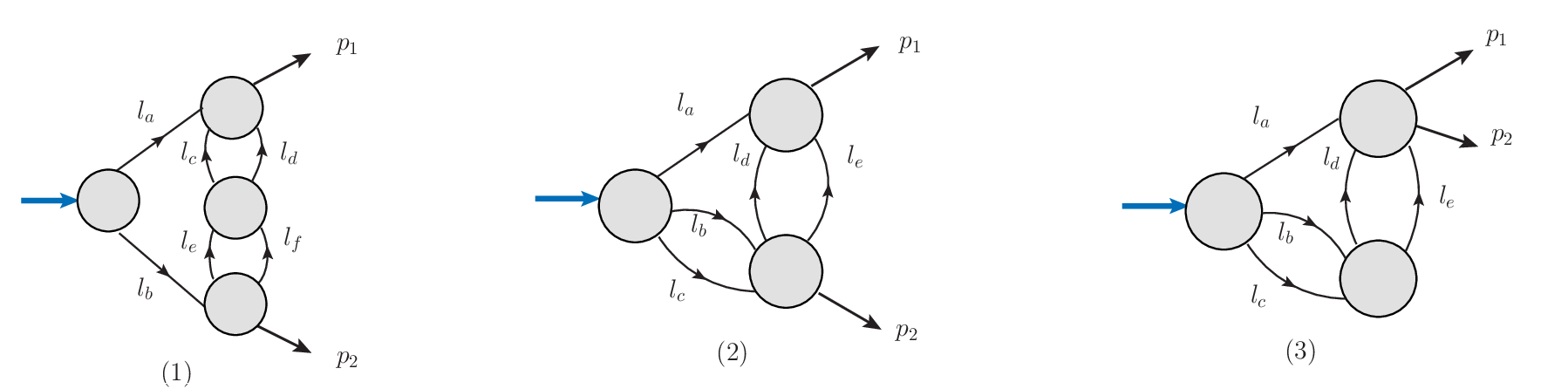} } 
	\caption{Planar ordering cuts for cut (1), (2) or (3).} 
	\label{fig:cut_123_order}
\end{figure}

\begin{figure}[t]
	\centerline{\includegraphics[height=8.3cm]{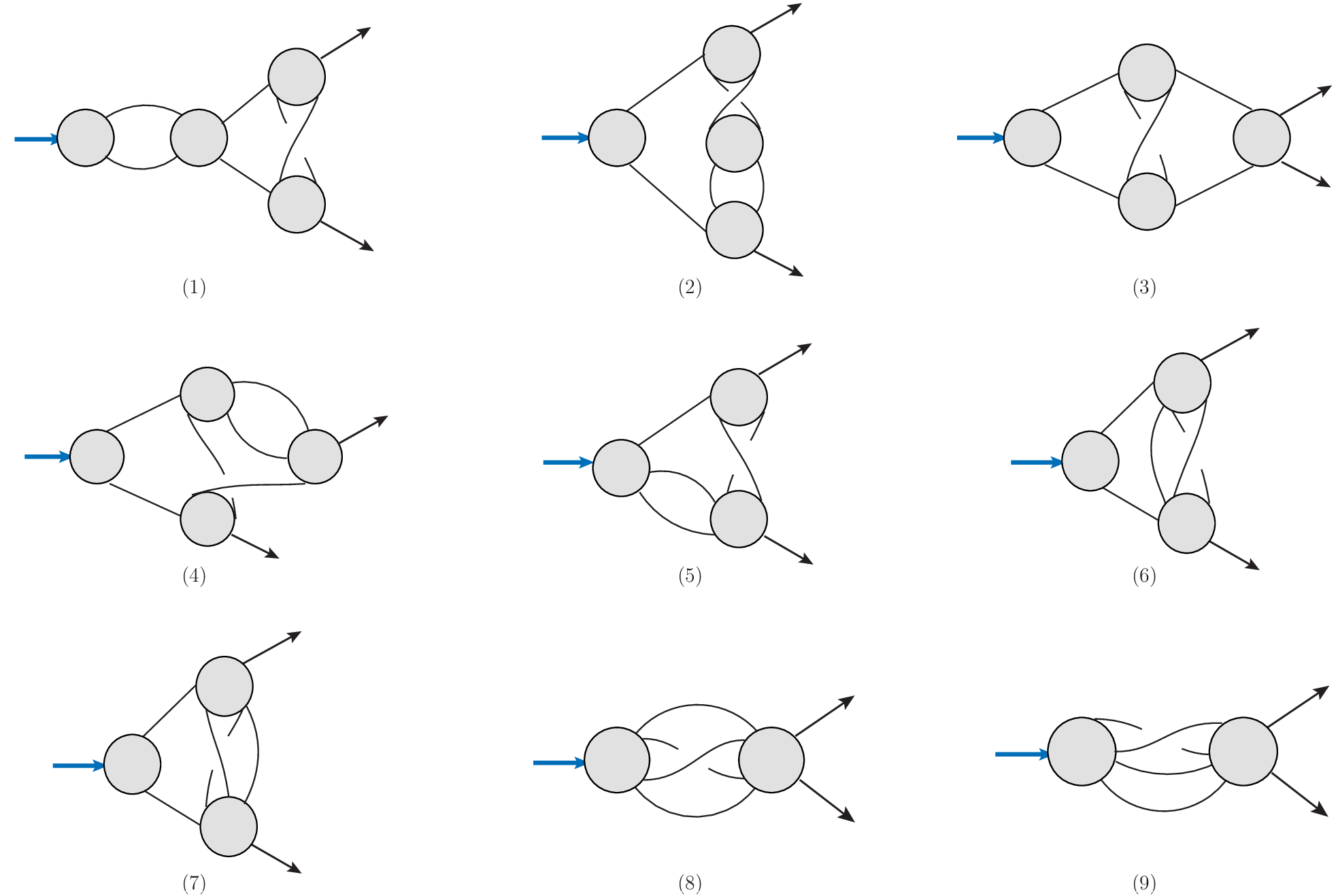} } 
	\caption{Some non-planar cuts we checked.} 
	\label{fig:3_loop_non_planar_cuts}
\end{figure}

Now we are ready to check our ansatz (with 202 free parameters) against the three failed cuts. As in the previous case, we start with the planar cuts, as shown in Figure~\ref{fig:cut_123_order}. We find that the fully deformed numerators $N_i^{[2]}$ can indeed pass all these cuts, and the number of parameters in $\Delta_{i}^{[2]}$ is reduced to 151.
We have also performed other checks including the non-planar cuts for $N_i$. 
We list these non-planar cuts in Figure~\ref{fig:3_loop_non_planar_cuts}.
Nicely, all other cuts are automatically satisfied.

Up to now, we have successfully obtained the full set of $N_i^{[1]}$ and $N_i^{[2]}$, both of which can pass all unitarity cuts. Thus $N_i = N_i^{[1]} + N_i^{[2]}$ is a physical solution for the three-loop Sudakov form factor in pure YM theory. Within the solution space, there are 2918 free parameters in total, 742 for $N_i^{[1]}$ and 2176 for $N_i^{[2]}$.
Due to the introduction of deformation, it is clear that our solution does not satisfy the global off-shell dual Jacobi relations.
However, one can show that the $N_i$'s we have obtained satisfy all ``on-shell" dual Jacobi relations, which is the topic of the next subsection.

\subsection{On-shell dual Jacobi relations}\label{sec:onshellJacobi}

The CK duality has been extremely useful in constructing the full-color integrand by reducing ansatz to a very small number of master numerators, as we have seen in the previous sections. Another important motivation for finding CK-dual solutions is related to the double-copy property: given a gauge-theory amplitude satisfying CK duality, it is straightforward to get gravitational amplitude by a double-copy of the kinematic numerators. 
However, it should be noted that to perform double-copy, it is sufficient to ask the CK duality to hold only at the unitarity-cut level, instead of globally on the off-shell loop integrand
\cite{Bern:2015ooa}.

\begin{figure}[t]
	\centerline{\includegraphics[height=5cm]{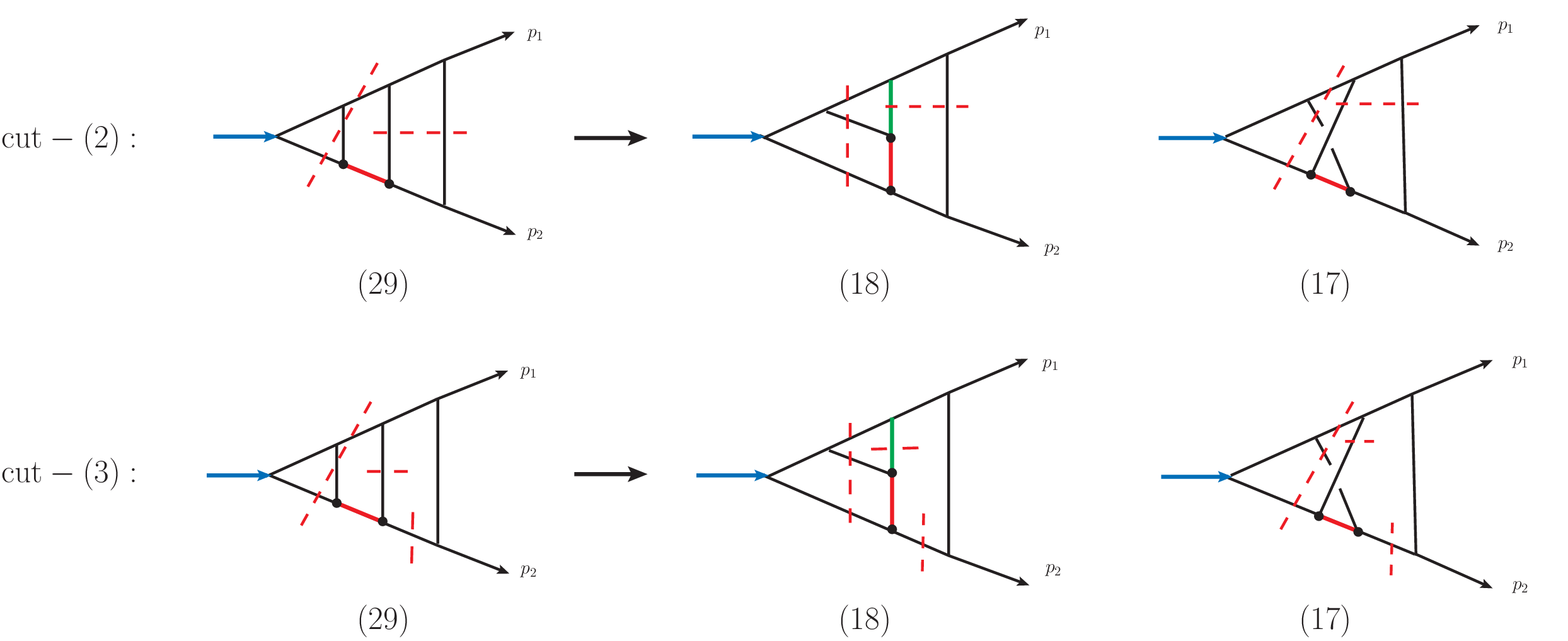} } 
	\caption{The Jacobi relation that we excluded in Fig \ref{fig:3_loop_ck_example2} contains 2 possible cuts that contribute to cuts (2) and (3). The Red line denotes the Jacobi propagator and the green one denotes the propagator that $\Delta_{18}^{[2]}$ always be proportional to. In both cases $\Delta_{18}^{[2]}$ will be cut to 0 and make the CK relation on-shell satisfied.} 
	\label{fig:tp_18_cut}
\end{figure}

Let us clarify more about the meaning of ``on-shell" CK relations.
Given a dual kinematic relation for three numerators, say $N_s + N_t + N_u = 0$, the on-shell version means that the relation should be true under all possible cuts that are shared by the three $s,t,u$ topologies, together with the condition that the Jacobi propagators $s,t,u$ are not cut. We will see concrete examples below as given in Figure~\ref{fig:tp_18_cut}.

Consider our solution $N_i = n_i + \Delta_i$, the global duality does not hold due to the deformation. Below we will show that a solution satisfying all on-shell dual Jacobi relations can be obtained. For this, we only need to check the on-shell dual relations for the deformation $\Delta_i$, since by construction the $n_i$ part already satisfies global CK relations. 
Furthermore, we only need to check on-shell CK relations within the cuts (1), (2), and (3) because $\Delta_i$'s vanish on other cuts, which makes the on-shell Jacobi relations trivial.

We summarize the steps of checking each dual Jacobi relation as follows:
\begin{enumerate}
	\item Check whether it is satisfied directly at the off-shell level.
	\item If not, identify all possible cuts for the three $s,t,u$ topologies that can be embedded into the cut (1), (2), and (3) while ensuring that the Jacobi propagators are not cut.
	\item For all the cuts found in the last step, check the Jacobi relations by imposing the on-shell cut conditions.
	\item If any of the on-shell dual Jacobi relations is not valid, use it as a constraint to reduce the solution space.
\end{enumerate}
Below we check $\Delta_i^{[1]}$ and $\Delta_i^{[2]}$ separately following the above steps.

First, for $\Delta_i^{[1]}$, we find all dual Jacobi relations among them are off-shell satisfied.
Second, for $\Delta_i^{[2]}$, not all the Jacobi relations among them are directly off-shell satisfied. 

As a concrete example, let us review the dual Jacobi relation \eqref{eq:delta29Jacobi} which we abandoned to prevent $\Delta_{29}^{[2]}$ and $\Delta_{45}^{[2]}$ from affecting cut (8). The solution with $\Delta_{29}^{[2]}=0$ clearly breaks this CK relation at off-shell level:
\begin{equation}
	\Delta_{29}^{[2]} + \Delta_{18}^{[2]} + \Delta_{17}^{[2]} = \Delta_{18}^{[2]} \neq 0\; ,
\end{equation}
since $\Delta_{17}^{[2]}=0=\Delta_{29}^{[2]}$ while $\Delta_{18}^{[2]}$ is non-zero. 
Next, we examine it at the on-shell level. This Jacobi relation possesses two kinds of cuts that contribute to cuts (2) and (3) respectively, and we illustrate them explicitly in Figure~\ref{fig:tp_18_cut}. We find that $\Delta_{18}^{[2]}$ is always proportional to the green-line propagator as shown in the figures, so in both cases $\Delta_{18}^{[2]}$ will be zero under the cut, which guarantees the Jacobi relation to be true at the on-shell level:
\begin{equation}\label{exclude_ck_on_shell_satisfy}
	(\Delta_{29}^{[2]} + \Delta_{18}^{[2]} + \Delta_{17}^{[2]}) \big|_{\textrm{cut-(2),(3)}} = \Delta_{18}^{[2]}\big|_{\textrm{cut-(2),(3)}} = 0 \,.
\end{equation}

\begin{figure}[t]
	\centerline{\includegraphics[height=2.3cm]{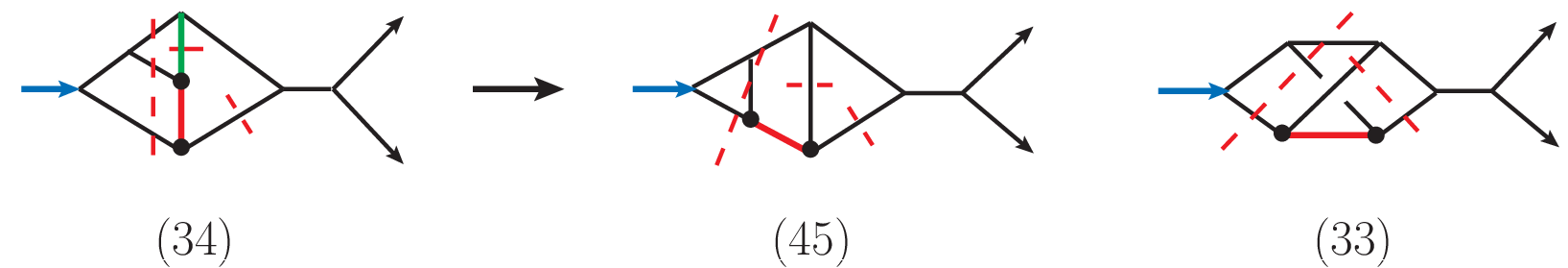} } 
	\caption{A dual Jacobi relation example that does not off-shell satisfy yet on-shell satisfy} 
	\label{fig:tp_34_CK}
\end{figure}

In a similar way, we check all the dual Jacobi relations among $\Delta_{i}^{[2]}$. Interestingly, it turns out that all of them are directly on-shell satisfied within the 151 parameters solution space. We display another example in Figure~\ref{fig:tp_34_CK}, this dual Jacobi relation also does not off-shell satisfy:
\begin{equation}
	\Delta_{34}^{[2]} + \Delta_{45}^{[2]} + \Delta_{33}^{[2]} = \Delta_{34}^{[2]} \neq 0 \,,
\end{equation}
because both $\Delta_{45}^{[2]}$ and $\Delta_{33}^{[2]}$ are 0, but $\Delta_{34}^{[2]}$ is non-zero. But if we consider it at the on-shell level, we find this Jacobi relation only possesses one cut contribution to cut (3) as shown in Figure~\ref{fig:tp_34_CK}. Under this cut, $\Delta_{34}^{[2]}$ will be cut to 0 because it is always proportional to the green-line propagator, which induces:
\begin{equation}
	(\Delta_{34}^{[2]} + \Delta_{45}^{[2]} + \Delta_{33}^{[2]}) \big|_{\textrm{cut-(3)}} = \Delta_{34}^{[2]} \big|_{\textrm{cut-(3)}} = 0 \,.
\end{equation}

Thus, we have checked that the numerator solution obtained in the previous subsection satisfies all dual Jacobi relations at the unitarity cut level. 
We would like to comment that for form factors, just satisfying dual Jacobi relations is not sufficient to apply double copy. This is because there are other operator-induced color relations that are important to consider to ensure diffeomorphism invariance for gravitational quantities, see \cite{Lin:2021pne, Lin:2022jrp, Lin:2023rwe} for discussion. We will not consider the double copy in this paper and leave it for future studies.

\section{Summary and discussion}
\label{sec:discussion}
\begin{figure}[t]
		\centerline{\includegraphics[height=9cm]{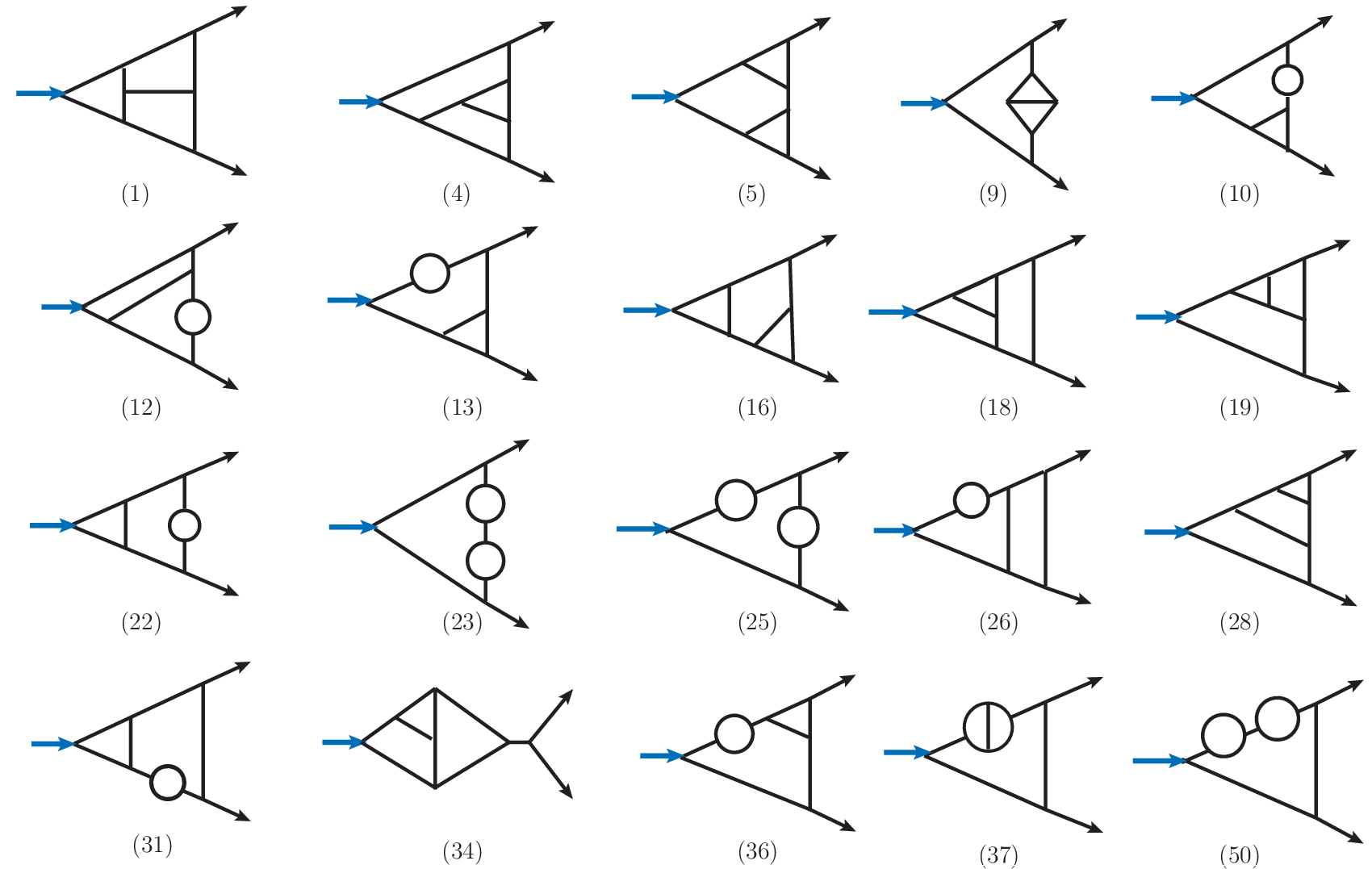} } 
		\caption{Topologies that have non-zero deformation.} 
		\label{fig:3_loop_deformed_topologies_2_non_zero}
\end{figure}

In this paper, we apply CK-duality to pure YM theory at the three-loop order, employing a strategy of minimal deformation that maximizes the use of CK relations. 
We obtain a compact integrand for the three-loop Sudakov form factor, with the final numerators expressed as
\begin{equation}
	N_i = \left\{ 
	\begin{matrix} 
	& n_i +\Delta_i, & \qquad\qquad i \in \{\textrm{topologies in Figure~\ref{fig:3_loop_deformed_topologies_2_non_zero}}\},  \\ 
	& n_i , & \textrm{other topologies}. 
	\end{matrix} \right.
\end{equation} 
Here, $n_i$'s are a set of numerators for Figure~\ref{fig:3_loop_all_topologies} that satisfy global dual-Jacobi relations, while $\Delta_i$'s denote the deformation which applies only to 20 planar topologies depicted in Figure~\ref{fig:3_loop_deformed_topologies_2_non_zero}.

To determine the complete set of $n_i$, three master numerators are needed.
Notably, for the deformation, only one master numerator for the formation $\Delta_{1}$ is necessary. 
Furthermore, the solution can be presented in a highly compact form as
\begin{equation}
	\Delta_{1} = \Delta_{1}^{[1]} + \Delta_{1}^{[2]}   \,,
\end{equation}
where (we present a specially simple solution for $\Delta_{1}$ in the solution space)
\begin{align}
\label{eq:delta11}
		\Delta_{1}^{[1]} & = \, 2(d-2)^2 (\varepsilon_1 \cdot \varepsilon_2)(p_1 \cdot p_2 ) \, l_2^2 \, l_3^2 \, l_4^2  \,, \\
		\Delta_{1}^{[2]} & = -2(d-2)^2 l_4^2 \Big(24 S_1^{[2]} + S_2^{[2]}- 5S_3^{[2]}+ 4 S_4^{[2]}- 2S_5^{[2]} -9S_6^{[2]} -15S_7^{[2]} -4S_8^{[2]} -20S_9^{[2]} \notag\\ \notag
		&-4S_{10}^{[2]} - 20S_{11}^{[2]} +4S_{12}^{[2]} + 12S_{13}^{[2]} +4S_{14}^{[2]} + 8S_{15}^{[2]} -24S_{16}^{[2]} - 4S_{17}^{[2]} 
		+4S_{18}^{[2]} + 28S_{19}^{[2]}+4S_{20}^{[2]} \\ \notag
		&- 20S_{21}^{[2]}
		+ 8S_{22}^{[2]}- 4S_{23}^{[2]}- 28S_{24}^{[2]} -40S_{25}^{[2]}- 4S_{26}^{[2]}- 8S_{27}^{[2]} +4S_{28}^{[2]}- 8S_{29}^{[2]} 
		+ 8S_{30}^{[2]} +8S_{31}^{[2]} \Big) \, ,
\end{align}
and $S_i^{[2]}$ are simple symmetric Lorentz product basis which are given in Appendix~\ref{app:symbasis}.

We strongly suggest interested readers to compare the expressions of $n_1$ and $\Delta_1$ to appreciate the simplicity of the latter. The above numerator solution, along with the relevant propagator lists and dual Jacobi relations, is given in the ancillary files.
As demonstrated, the solution also satisfies all dual Jacobi relations at the unitarity cut level.

\begin{figure}[t]
	\centerline{\includegraphics[height=2.8cm]{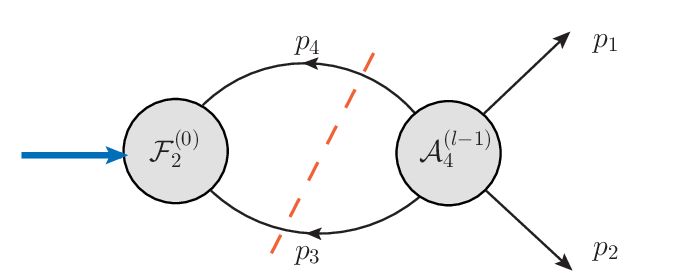} } 
	\caption{The $l$-loop Sudakov form factor and $(l-1)$-loop four-point amplitude are related by a double cut.} 
	\label{fig:double_cut_for_FF_AM}
\end{figure}

In the remainder of this section, we discuss the relation between the Sudakov form factor and the four-point amplitudes.
Sudakov form factor and four-point amplitude are closely related by the double-cut illustrated in Figure~\ref{fig:double_cut_for_FF_AM}, where the $(l-1)$-loop four-point amplitude will appear as a sub-block for the $l$-loop Sudakov form factor. Naturally, we expect that the three-loop Sudakov form factor will inherit the property of two-loop four-point amplitude through this cut. 

\begin{figure}[t]
	\centerline{\includegraphics[height=2.8cm]{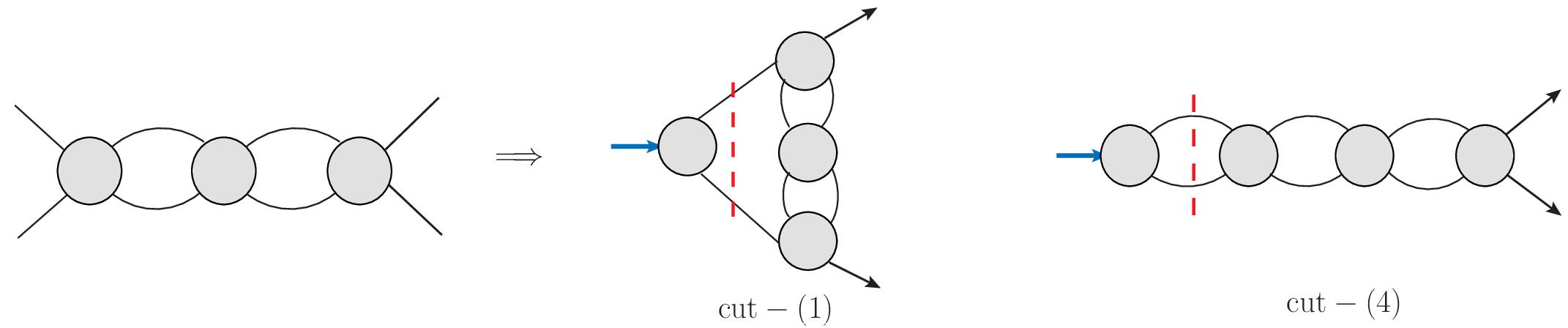} } 
	\caption{The deformed cut for four-point two-loop amplitude will imply two possible cuts for the three-loop Sudakov form factor. } 
	\label{fig:FF_Am_Deformation_Compare}
\end{figure}

For the four-point two-loop amplitude, it was shown in \cite{Li:2023akg} that the deformation can be made to the two double-cut shown on the left-hand side of  Figure~\ref{fig:FF_Am_Deformation_Compare}.
This is supposed to induce deformation for the cuts (1) and (4) for the three-loop form factor, both of which contain the deformed cut of the four-point amplitude. 
However, as we have seen in previous sections, only the cut (1) requires deformation, while the cut (4) can be satisfied directly by the global CK-dual integrand (\emph{i.e.}, the $n_i$ part). 

\begin{figure}[t]
	\centerline{\includegraphics[height=2.8cm]{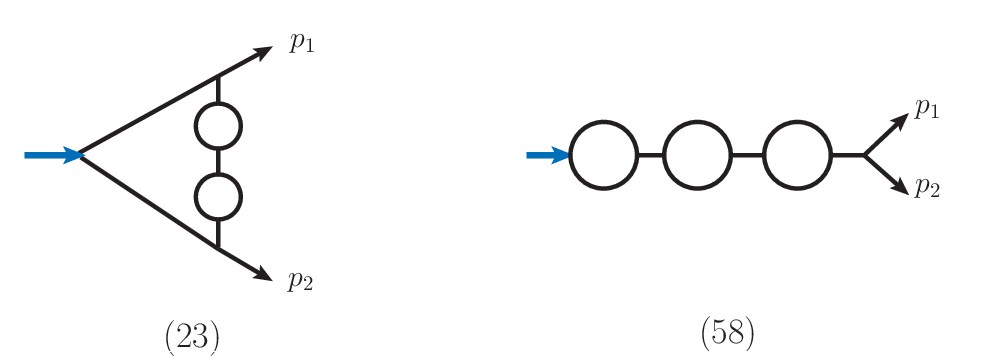} } 
	\caption{Topology (23) and (58) which contribute to cuts (1) and (4) respectively. Topology (23) possess an irreducible numerator while topology (58) does not.} 
	\label{fig:Double_bubble}
\end{figure}

This difference is not due to any inconsistency but reveals important different properties between the Sudakov form factor and four-point amplitude.
First, the symmetry properties of the two quantities are different.
For example, for the amplitude, the master double-box topology has both vertical and horizontal flip symmetry. On the other hand, after sewing two external legs with the form factor vertex, the three-loop form factor topologies, such as (1) and (29) in Figure~\ref{fig:3_loop_all_topologies},  will only possess half of the symmetry of the amplitude diagram. 
Second, the tree products of cuts (1) and (4) have different properties for the numerators of maximal topologies. 
For example, consider the two form factor topologies (23) and (58) as  shown in Figure \ref{fig:Double_bubble}.
The topology (23) contains an irreducible numerator,
while the numerator of topology (58) is always proportional to $p_1 \cdot p_2$ which can shrink one of the propagators. 
This implies that the contribution of topology (58) can be redistributed into other topologies. In this sense, the CK-dual ansatz is easier to satisfy cut (4) because of the freedom of redistribution.

Another difference to note is that cuts (2) and (3) can not be embedded into the double-cut in Figure~\ref{fig:double_cut_for_FF_AM}. This implies that the deformation for the three-loop form factor can not be fully determined by the lower loop four-point amplitude, new deformation structures will appear as the number of loops increases.

\begin{figure}[t]
	\centerline{\includegraphics[height=2.5cm]{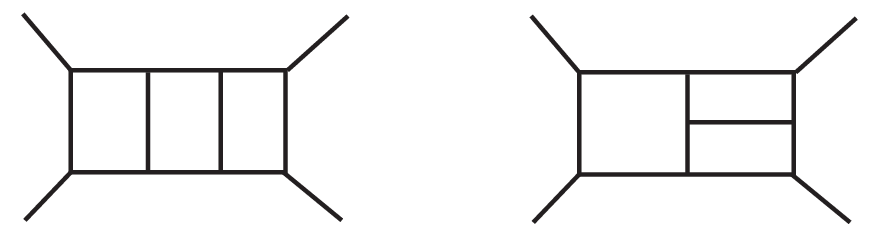} } 
	\caption{Possible master deformation topology for four-point three-loop amplitude.} 
	\label{fig:3_loop_Am_Deform_Master}
\end{figure}

It would be highly interesting to generalize the above consideration to the three-loop four-point amplitude in pure YM.
Based on the result of the four-point two-loop amplitude and three-loop Sudakov form factor, it is natural to expect that both the two topologies in Figure~\ref{fig:3_loop_Am_Deform_Master} need deformation and they should both serve as master topologies for the deformation.
Moreover, the three-loop form factor result also implies that multiple cuts for the three-loop amplitude, similar to that of cuts (1), (2), and (3), require deformation. 
We leave the detailed construction of the three-loop amplitude to another work.

\acknowledgments

This work is supported in part by the National Natural Science Foundation of China (Grants No.~12175291, 11935013, 12047503), and the Chinese Academy of Sciences (Grant No. YSBR-101). 
We also acknowledge the support of the HPC Cluster of ITP-CAS.

\appendix

\section{Complete CK relations for three-loop Sudakov form factor}\label{app:three_loop_CKrelation}
In this appendix, we give the set of dual Jacobi relations for obtaining $n_{4}, \ldots, n_{58}$ from the three master numerators:
\begin{align}
	n_{4} &= -n_{1}[p_1, p_2, l_1, -l_1 + l_2, l_3] + n_{2}[p_2, p_1, l_1, l_1 - l_2 - p_1, l_1 - l_2 - l_3 - p_1 - p_2] \notag \\ 
	n_{5} &= n_{1} - n_{2}[p_1, p_2, -l_1 + p_1 + p_2, -l_1 + l_3 + p_1 + p_2, -l_1 - l_2 + l_3 + p_2] \notag \\ 
	n_{6} &= n_{2}[p_1, p_2, l_1, l_1 - l_3, l_1 + l_2] + n_{3}[p_1, p_2, l_1, -l_1 - l_2, -l_3 + p_2] \notag \\ 
	n_{7} &= n_{2}[p_1, p_2, l_1, l_1 - l_2, l_3] - n_{6}[p_2, p_1, -l_1 + p_1 + p_2, l_2 + l_3 - p_2, -l_3] \notag \\ 
	n_{8} &= -n_{3}[p_1, p_2, l_1, -l_3, l_2 + l_3 + p_2] - n_{3}[p_1, p_2, -l_1 + p_1 + p_2, -l_2 - l_3, l_3] \notag \\ 
	n_{9} &= -n_{4}[p_1, p_2, l_1, l_2, l_1 + l_3 - p_1 - p_2] - n_{4}[p_1, p_2, l_1, l_1 - l_2 - p_1, -l_3 - p_2] \notag \\ 
	n_{10} &= -n_{4} - n_{4}[p_1, p_2, l_1, l_1 - l_2 - p_1, l_3] \notag \\ 
	n_{11} &= n_{4}[p_1, p_2, l_1, -l_2 - l_3, l_1 + l_2 - p_1 - p_2] + n_{4}[p_1, p_2, l_1, l_3, l_1 + l_2 - p_1 - p_2] \notag \\ 
	n_{12} &= n_{4}[p_1, p_2, l_1, l_2 - l_3, l_1 - l_2 - p_1 - p_2] + n_{4}[p_1, p_2, l_1, l_1 + l_3 - p_1, l_1 - l_2 - p_1 - p_2] \notag \\ 
	n_{13} &= n_{5}[p_2, p_1, -l_1 + p_1 + p_2, l_3, -l_1 - l_2] + n_{5} \notag \\ 
	n_{14} &= -n_{6}[p_1, p_2, l_1, -l_1 - l_2, -l_3 + p_2] + n_{6} \notag \\ 
	n_{15} &= -n_{6} - n_{6}[p_2, p_1, l_1, l_2, -l_2 - l_3] \notag \\ 
	n_{16} &= n_{1} - n_{6}[p_1, p_2, -l_1 + p_1 + p_2, -l_2 - l_3 - p_1 - p_2, l_3 + p_2] \notag \\ 
	n_{17} &= n_{6}[p_1, p_2, l_1, l_2, -l_2 - l_3 - p_1] + n_{6}[p_1, p_2, -l_1 + p_1 + p_2, l_3, -l_2 - l_3 - p_1] \notag \\ 
	n_{18} &= -n_{1}[p_1, p_2, l_1, l_2, -l_2 - l_3 - p_1 - p_2] + n_{6}[p_2, p_1, l_1, l_2, l_3 + p_1] \notag \\ 
	n_{19} &= -n_{7}[p_1, p_2, -l_1 + p_1 + p_2, -l_1 - l_2 - l_3 + p_2, l_3] - n_{18} \notag \\ 
	n_{20} &= -n_{7}[p_1, p_2, l_1, l_2, l_1 - l_2 - l_3 - p_1] - n_{7}[p_2, p_1, l_1, l_1 - l_2, l_2 - l_3 - p_2] \notag \\ 
	n_{21} &= -n_{8}[p_1, p_2, l_1, -l_2, l_3] + n_{11}[p_2, p_1, -l_1 + p_1 + p_2, l_2, -l_3] \notag \\ 
	n_{22} &= n_{8}[p_2, p_1, -l_1 + p_1 + p_2, -l_2 - p_1, l_3] - n_{12}[p_2, p_1, -l_1 + p_1 + p_2, -l_1 - l_2, -l_3] \notag \\ 
	n_{23} &= n_{9} + n_{9}[p_1, p_2, l_1, l_1 - l_2 - p_1, l_3] \notag \\ 
	n_{24} &= n_{9}[p_1, p_2, l_1, -l_3, -l_1 + l_2 + p_1] + n_{9}[p_1, p_2, l_1, l_1 - l_2 + l_3 - p_1, -l_1 + l_2 + p_1] \notag \\ 
	n_{25} &= n_{10}[p_2, p_1, -l_1 + p_1 + p_2, l_3, -l_1 - l_2] + n_{10}[p_2, p_1, -l_1 + p_1 + p_2, l_3, l_2] \notag \\ 
	n_{26} &= n_{13} - n_{20}[p_1, p_2, -l_1 + p_1 + p_2, -l_3, l_2] \notag \\ 
	n_{27} &= n_{13} + n_{13}[p_1, p_2, l_1, l_2, l_1 - l_3 - p_1 - p_2] \notag \\ 
	n_{28} &= n_{4} - n_{14}[p_1, p_2, -l_1 + p_1 + p_2, l_2 + l_3, -l_3] \notag \\ 
	n_{29} &= n_{15}[p_2, p_1, l_1, l_2, l_3 - p_2] + n_{16}[p_2, p_1, l_1, l_2, l_3 - p_1 - p_2] \notag \\ 
	n_{30} &= -n_{15}[p_1, p_2, l_1, l_3, l_2] + n_{15}[p_1, p_2, -l_1 + p_1 + p_2, l_3, l_2] \notag \\ 
	n_{31} &= n_{16} + n_{16}[p_1, p_2, l_1, l_2, -l_2 - l_3 - p_1 - p_2] \notag \\ 
	n_{32} &= n_{16} - n_{16}[p_1, p_2, -l_1 + p_1 + p_2, l_2, l_3] \notag \\ 
	n_{33} &= n_{17} - n_{17}[p_2, p_1, l_1, l_2, l_3] \notag \\ 
	n_{34} &= n_{18} - n_{18}[p_2, p_1, l_1, l_2, l_3] \notag \\ 
	n_{35} &= -n_{19}[p_2, p_1, -l_1 + p_1 + p_2, l_3, l_1 - l_2 - l_3 - p_1 - p_2] \notag \\
						 &\quad - n_{19}[p_2, p_1, -l_1 + p_1 + p_2, l_1 - l_2 - l_3 - p_1 - p_2, l_3] \notag \\ 
	n_{36} &= -n_{19}[p_1, p_2, l_1, -l_1 - l_2, l_2 - l_3] - n_{19}[p_1, p_2, l_1, l_2, -l_1 - l_2 - l_3] \notag \\ 
	n_{37} &= -n_{19}[p_1, p_2, l_1, -l_1 - l_2, -l_3] - n_{19} \notag \\ 
	n_{38} &= n_{21} - n_{21}[p_1, p_2, -l_1 + p_1 + p_2, -l_2, -l_3] \notag \\ 
	n_{39} &= n_{22} - n_{22}[p_1, p_2, -l_1 + p_1 + p_2, l_2, l_3] \notag \\ 
	n_{40} &= -n_{26}[p_1, p_2, -l_1 + p_1 + p_2, l_2, l_3] + n_{26}[p_2, p_1, -l_1 + p_1 + p_2, l_2, -l_3 - p_1 - p_2] \notag \\ 
	n_{41} &= -n_{26}[p_1, p_2, l_1, l_2, l_1 + l_3 - p_1 - p_2] + n_{26}[p_2, p_1, l_1, l_2, l_1 + l_3 - p_1 - p_2] \notag \\ 
	n_{42} &= n_{27}[p_2, p_1, l_1, l_2, -l_3] - n_{27}[p_1, p_2, l_1, l_2, -l_3] \notag \\ 
	n_{43} &= -n_{29} + n_{29}[p_1, p_2, -l_1 + p_1 + p_2, l_2, l_3] \notag \\ 
	n_{44} &= n_{29}[p_1, p_2, l_1, l_2, -l_3] - n_{29}[p_1, p_2, -l_1 + p_1 + p_2, -l_2 - p_1 - p_2, -l_3] \notag \\ 
	n_{45} &= n_{29}[p_1, p_2, l_1, l_2, -l_2 - l_3] - n_{29}[p_2, p_1, l_1, l_2, -l_2 - l_3] \notag \\ 
	n_{46} &= -n_{31}[p_2, p_1, -l_1 + p_1 + p_2, -l_2 - p_1 - p_2, -l_3] + n_{31}[p_1, p_2, -l_1 + p_1 + p_2, -l_2 - p_1 - p_2, -l_3] \notag \\ 
	n_{47} &= n_{31}[p_2, p_1, l_1, -l_2 - p_1 - p_2, -l_3] - n_{31}[p_2, p_1, -l_1 + p_1 + p_2, -l_2 - p_1 - p_2, -l_3] \notag \\ 
	n_{48} &= -n_{34}[p_1, p_2, l_1, -l_1 - l_2, l_2 + l_3] - n_{34}[p_1, p_2, l_1, l_2, -l_2 - l_3] \notag \\ 
	n_{49} &= n_{11} + n_{35}[p_1, p_2, l_1, l_1 + l_2 - p_1 - p_2, l_3] \notag \\ 
	n_{50} &= n_{36}[p_1, p_2, l_1, l_2, -l_3] + n_{36}[p_1, p_2, l_1, l_2, -l_1 + l_3] \notag \\ 
	n_{51} &= n_{38}[p_1, p_2, -l_1 + p_1 + p_2, l_2, l_3] - n_{46}[p_1, p_2, l_1, l_2, -l_3] \notag \\ 
	n_{52} &= n_{40}[p_1, p_2, l_1, l_2, -l_3] + n_{40}[p_1, p_2, l_1, l_2, l_3 - p_1 - p_2] \notag \\ 
	n_{53} &= -n_{41}[p_1, p_2, l_1, l_2, -l_1 - l_3] - n_{41} \notag \\ 
	n_{54} &= n_{43}[p_1, p_2, l_1, l_2, -l_3] + n_{43}[p_1, p_2, l_1, -l_2 - p_1 - p_2, -l_3] \notag \\ 
	n_{55} &= n_{43}[p_1, p_2, l_1, l_2, -l_2 - l_3] - n_{43}[p_2, p_1, l_1, l_2, -l_2 - l_3] \notag \\ 
	n_{56} &= n_{44}[p_1, p_2, l_1, l_2, -l_3] + n_{44}[p_1, p_2, l_1, l_2, l_3 - p_1 - p_2] \notag \\ 
	n_{57} &= n_{46} - n_{46}[p_1, p_2, -l_1 + p_1 + p_2, l_2, l_3] \notag \\ 
	n_{58} &= n_{54}[p_1, p_2, l_1, l_2, -l_3] + n_{54}[p_1, p_2, l_1, l_2, l_3 - p_1 - p_2] \,.
\end{align}
Corresponding figures and labeling of momenta are given in Figure~\ref{fig:3_loop_all_topologies}.

\section{Symmetry basis}\label{app:symbasis}
In this appendix, we provide the explicit form for the Lorentz product basis $S_i^{[2]}$ in \eqref{eq:delta11}:
\begin{align}
	S_1^{[2]} &=  (\varepsilon_1 \cdot l_8)(\varepsilon_2 \cdot l_7)l_2^2 l_3^2 \,, \qquad 
	S_2^{[2]} =  (\varepsilon_1 \cdot p_2)(\varepsilon_2 \cdot p_1)l_2^2 l_3^2 \\ \notag
	S_3^{[2]} &=  (\varepsilon_1 \cdot l_3)(\varepsilon_2 \cdot l_3)(l_2^2)^2 + (\varepsilon_1 \cdot l_2)(\varepsilon_2 \cdot l_2)(l_3^2)^2 \\ \notag
	S_4^{[2]} &=  (\varepsilon_1 \cdot l_3)(\varepsilon_2 \cdot l_2)(l_2^2)^2 + (\varepsilon_1 \cdot l_3)(\varepsilon_2 \cdot l_2)(l_3^2)^2 \\ \notag
	S_5^{[2]} &=  (\varepsilon_1 \cdot l_3)(\varepsilon_2 \cdot l_4)(l_2^2)^2 + (\varepsilon_1 \cdot l_4)(\varepsilon_2 \cdot l_2)(l_3^2)^2 \\ \notag
	S_6^{[2]} &=  (\varepsilon_1 \cdot l_2)(\varepsilon_2 \cdot l_3)(l_2^2)^2 + (\varepsilon_1 \cdot l_2)(\varepsilon_2 \cdot l_3)(l_3^2)^2 \\ \notag
	S_7^{[2]} &=  (\varepsilon_1 \cdot l_4)(\varepsilon_2 \cdot l_3)(l_2^2)^2 + (\varepsilon_1 \cdot l_2)(\varepsilon_2 \cdot l_4)(l_3^2)^2 \\ \notag
	S_8^{[2]} &=  (\varepsilon_1 \cdot l_9)(\varepsilon_2 \cdot l_2)l_2^2 (l_2 \cdot l_3) + (\varepsilon_1 \cdot l_3)(\varepsilon_2 \cdot l_1)l_3^2 (l_2 \cdot l_3) \\ \notag
	S_{9}^{[2]} &=  (\varepsilon_1 \cdot l_3)(\varepsilon_2 \cdot l_3)l_2^2 (l_2 \cdot l_3) + (\varepsilon_1 \cdot l_2)(\varepsilon_2 \cdot l_2)l_3^2 (l_2 \cdot l_3) \\ \notag
	S_{10}^{[2]} &=  (\varepsilon_1 \cdot l_3)(\varepsilon_2 \cdot l_2)l_2^2 (l_2 \cdot l_3) + (\varepsilon_1 \cdot l_3)(\varepsilon_2 \cdot l_2)l_3^2 (l_2 \cdot l_3) \\ \notag
	S_{11}^{[2]} &=  (\varepsilon_1 \cdot l_3)(\varepsilon_2 \cdot l_4)l_2^2 (l_2 \cdot l_3) + (\varepsilon_1 \cdot l_4)(\varepsilon_2 \cdot l_2)l_3^2 (l_2 \cdot l_3) \\ \notag
	S_{12}^{[2]} &=  (\varepsilon_1 \cdot l_2)(\varepsilon_2 \cdot l_9)l_2^2 (l_2 \cdot l_3) + (\varepsilon_1 \cdot l_1)(\varepsilon_2 \cdot l_3)l_3^2 (l_2 \cdot l_3) \\ \notag
	S_{13}^{[2]} &=  (\varepsilon_1 \cdot l_2)(\varepsilon_2 \cdot l_3)l_2^2 (l_2 \cdot l_3) + (\varepsilon_1 \cdot l_2)(\varepsilon_2 \cdot l_3)l_3^2 (l_2 \cdot l_3) \\ \notag
	S_{14}^{[2]} &=  (\varepsilon_1 \cdot l_2)(\varepsilon_2 \cdot l_2)l_2^2 (l_2 \cdot l_3) + (\varepsilon_1 \cdot l_3)(\varepsilon_2 \cdot l_3)l_3^2 (l_2 \cdot l_3) \\ \notag
	S_{15}^{[2]} &=  (\varepsilon_1 \cdot l_2)(\varepsilon_2 \cdot l_4)l_2^2 (l_2 \cdot l_3) + (\varepsilon_1 \cdot l_4)(\varepsilon_2 \cdot l_3)l_3^2 (l_2 \cdot l_3) \\ \notag
	S_{16}^{[2]} &=  (\varepsilon_1 \cdot l_4)(\varepsilon_2 \cdot l_3)l_2^2 (l_2 \cdot l_3) + (\varepsilon_1 \cdot l_2)(\varepsilon_2 \cdot l_4)l_3^2 (l_2 \cdot l_3) \\ \notag
	S_{17}^{[2]} &=  (\varepsilon_1 \cdot l_4)(\varepsilon_2 \cdot l_2)l_2^2 (l_2 \cdot l_3) + (\varepsilon_1 \cdot l_3)(\varepsilon_2 \cdot l_4)l_3^2 (l_2 \cdot l_3) \\ \notag
	S_{18}^{[2]} &=  (\varepsilon_1 \cdot l_2)(\varepsilon_2 \cdot l_2)l_2^2 (l_3 \cdot l_4) + (\varepsilon_1 \cdot l_3)(\varepsilon_2 \cdot l_3)l_3^2 (l_2 \cdot l_4) \\ \notag
	S_{19}^{[2]} &=  (\varepsilon_1 \cdot l_2)(\varepsilon_2 \cdot l_3)l_2^2 (l_3 \cdot l_4) + (\varepsilon_1 \cdot l_2)(\varepsilon_2 \cdot l_3)l_3^2 (l_2 \cdot l_4) \\ \notag
	S_{20}^{[2]} &= (\varepsilon_1 \cdot l_2)(\varepsilon_2 \cdot l_4)l_2^2 (l_3 \cdot l_4) + (\varepsilon_1 \cdot l_4)(\varepsilon_2 \cdot l_3)l_3^2 (l_2 \cdot l_4) \\ \notag
	S_{21}^{[2]} &= (\varepsilon_1 \cdot l_3)(\varepsilon_2 \cdot l_3)l_2^2 (l_2 \cdot l_4) + (\varepsilon_1 \cdot l_2)(\varepsilon_2 \cdot l_2)l_3^2 (l_3 \cdot l_4) \\ \notag
	S_{22}^{[2]} &=  (\varepsilon_1 \cdot l_3)(\varepsilon_2 \cdot l_2)l_2^2 (l_2 \cdot l_4) + (\varepsilon_1 \cdot l_3)(\varepsilon_2 \cdot l_2)l_3^2 (l_3 \cdot l_4) \\ \notag
	S_{23}^{[2]} &=   (\varepsilon_1 \cdot l_3)(\varepsilon_2 \cdot l_4)l_2^2 (l_2 \cdot l_4) + (\varepsilon_1 \cdot l_4)(\varepsilon_2 \cdot l_2)l_3^2 (l_3 \cdot l_4) \\ \notag
	S_{24}^{[2]} &= (\varepsilon_1 \cdot l_2)(\varepsilon_2 \cdot l_3)l_2^2 (l_2 \cdot l_4) + (\varepsilon_1 \cdot l_2)(\varepsilon_2 \cdot l_3)l_3^2 (l_3 \cdot l_4) \\ \notag
	S_{25}^{[2]} &=  (\varepsilon_1 \cdot l_4)(\varepsilon_2 \cdot l_3)l_2^2 (l_2 \cdot l_4) + (\varepsilon_1 \cdot l_2)(\varepsilon_2 \cdot l_4)l_3^2 (l_3 \cdot l_4) \\ \notag
	S_{26}^{[2]} &=  (\varepsilon_1 \cdot l_2)(\varepsilon_2 \cdot l_1)l_2^2 (l_3 \cdot l_7) + (\varepsilon_1 \cdot l_9)(\varepsilon_2 \cdot l_3)l_3^2 (l_2 \cdot l_8) \\ \notag
	S_{27}^{[2]} &=   (\varepsilon_1 \cdot l_4)(\varepsilon_2 \cdot l_1)l_2^2 (l_3 \cdot l_7) + (\varepsilon_1 \cdot l_9)(\varepsilon_2 \cdot l_4)l_3^2 (l_2 \cdot l_8) \\ \notag
	S_{28}^{[2]} &=  (\varepsilon_1 \cdot l_1)(\varepsilon_2 \cdot l_2)l_2^2 (l_3 \cdot l_7) + (\varepsilon_1 \cdot l_3)(\varepsilon_2 \cdot l_9)l_3^2 (l_2 \cdot l_8) \\ \notag
	S_{29}^{[2]} &= (\varepsilon_1 \cdot l_4)(\varepsilon_2 \cdot l_3)l_2^2 (l_3 \cdot l_7) + (\varepsilon_1 \cdot l_2)(\varepsilon_2 \cdot l_4)l_3^2 (l_2 \cdot l_8) \\ \notag
	S_{30}^{[2]} &=  (\varepsilon_1 \cdot l_1)(\varepsilon_2 \cdot l_4)l_2^2 (l_3 \cdot l_7) + (\varepsilon_1 \cdot l_4)(\varepsilon_2 \cdot l_9)l_3^2 (l_2 \cdot l_8) \\ \notag
	S_{31}^{[2]} &=  (\varepsilon_1 \cdot l_3)(\varepsilon_2 \cdot l_4)l_2^2 (l_3 \cdot l_7) + (\varepsilon_1 \cdot l_4)(\varepsilon_2 \cdot l_2)l_3^2 (l_2 \cdot l_8) \,.\notag
\end{align}
where the labeling of momenta follows Figure~\ref{fig:3_loop_tp1_label_all}, and the basis preserves the symmetry of the topologies.

\providecommand{\href}[2]{#2}\begingroup\raggedright\endgroup

\end{document}